\documentclass[fleqn,usenatbib]{mnras}

\usepackage{newtxtext,newtxmath}

\usepackage[T1]{fontenc}

\DeclareRobustCommand{\VAN}[3]{#2}
\let\VANthebibliography\thebibliography
\def\thebibliography{\DeclareRobustCommand{\VAN}[3]{##3}\VANthebibliography}

\usepackage{graphicx}	
\usepackage{amsmath}	
\usepackage{multirow}
\usepackage{subfig}
\usepackage{comment}
\usepackage{xcolor}
\newcommand{\mbf}[1]{\mathbf{#1}}

\newcommand{\CBB}[1]{C^{BB,{#1}}_\ell}

\newcommand{\pc}[3]{${#1}^{+{#2}}_{-{#3}}$}

\newcommand{\cc}[1]{\left({#1}\right)}
\newcommand{\rr}[1]{\left[{#1}\right]}
\newcommand{\be}{\begin{equation}}
\newcommand{\ee}{\end{equation}}
\def\bear#1\ear{\begin{align}#1\end{align}}
\newcommand{\nline}{\notag \\}
\newcommand{\f}{\frac}
\newcommand{\de}{\mathrm{d}}

\newcommand{\e}{\mathrm{e}}

\renewcommand{\mathbf}[1]{\mbox{\boldmath $#1$}}

\title[Reionization bias of primordial GWs]{A framework to mitigate patchy reionization contamination on the primordial gravitational wave signal}

\author[Jain et al.]{
Divesh Jain$^{1}$\thanks{djain@ncra.tifr.res.in}, 
Tirthankar Roy Choudhury$^{1}$\thanks{tirth@ncra.tifr.res.in},
Suvodip Mukherjee$^{2}$\thanks{suvodip@tifr.res.in},
and Sourabh Paul$^{3}$\thanks{sourabh.paul2@mcgill.ca}
\\
$^{1}$ National Centre for Radio Astrophysics, Tata Institute of Fundamental Research, Pune 411007, India\\
$^{2}$ Department of Astronomy \& Astrophysics, Tata Institute of Fundamental Research, 1, Homi Bhabha Road, Colaba, Mumbai 400005, India\\
$^{3}$ Department of Physics, McGill University, Montreal, QC, Canada
}

\date{Accepted XXX. Received YYY; in original form ZZZ}

\pubyear{2015}

\begin{document}
\label{firstpage}
\pagerange{\pageref{firstpage}--\pageref{lastpage}}
\maketitle

\begin{abstract}
One of the major goals of future cosmic microwave background (CMB) $B$-mode polarization experiments is the detection of primordial gravitational waves through an unbiased measurement of the tensor-to-scalar ratio $r$. Robust detection of this signal will require mitigating all possible contamination to the $B$-mode polarization from astrophysical origins. One such extragalactic contamination arises from the patchiness in the electron density during the reionization epoch. Along with the signature on CMB polarization, the patchy reionization can source secondary anisotropies on the CMB temperature through the kinetic Sunyaev-Zeldovich (kSZ) effect. In order to study the impact of this foreground for the upcoming CMB missions, we present a self-consistent framework to compute the CMB anisotropies based on a physically motivated model of reionization. We show that the value of $r$ can bias towards a higher value if the secondary contribution from reionization is neglected. However, combining small-scale kSZ signal, large-scale $E$-mode polarization, and $B$-mode polarization measurements, we can put constraints on the patchiness in electron density during reionization and can mitigate its impact on the value of $r$. CMB missions such as CMB-S4 and PICO may experience a bias of $>0.17\sigma$ which can go as high as $\sim 0.73\sigma$ for extreme reionization models allowed by the Planck and SPT CMB measurements. As future experiments target to measure $r$ at $5\sigma$, this is likely to affect the measurement significance and hence possibly affect the claim of detection of $r$, if not mitigated properly by using joint estimations of different reionization observables.
\end{abstract}




\section{Introduction}

Detection of the primordial $B$-mode polarization signal from the cosmic microwave background (CMB) will be a cornerstone of our understanding of the primordial gravitational waves produced during inflation. For example, the amplitude of this signal will determine the energy scale of inflation as well as give us insights into the nature of the inflaton field \citep[][and references therein]{doi:10.1146/annurev-astro-081915-023433,guzzetti2016gravitational}. This amplitude is usually described in terms of the primordial tensor-to-scalar power spectrum ratio $r$ defined at a wave mode $k_0 = 0.05$~Mpc$^{-1}$. The latest constraint on $r$ was obtained by \cite{arxiv.2203.16556} and is $r<0.035$ at 95$\%$ confidence level. In the next couple of decades, CMB experiments like Simons Observatory \citep{Ade_2019}, LiteBIRD \citep{suzuki2018litebird}, CMB-S4 \citep{Abazajian2019}, PICO \citep{hanany2019pico} are planned to provide sensitive measurements of $B$-mode polarization at large angular scales and aim to make the first statistically significant constraint on $r$.

The primal challenge in this effort is to correctly factor in the $B$-mode polarization foregrounds, mostly due to the secondary $B$-mode polarization signals arising in the post-recombination epochs. There are three primary contributors to this foreground, the galactic $B$-mode foreground, the lensing $B$-mode foreground, and the patchy reionization $B$-mode foreground \citep{HU1997323,hu2000reionization,lewis2006weak,2009AIPC.1141..121S,10.1093/ptep/ptu065,doi:10.1146/annurev-astro-081915-023433,mukherjee2019patchy}. The polarized thermal emission from dust within our galaxy and galactic synchrotron contribute to the galactic component \citep{HU1997323,doi:10.1146/annurev-astro-081915-023433,2020A&A...641A..11P}. The gravitational lensing of the CMB by large-scale structures in the late Universe converts the $E$-mode signal to $B$-modes, which constitute the lensing foregrounds \citep{lewis2006weak,10.1093/ptep/ptu065}. Finally, the $B$-mode polarization arising from the patchiness in the reionization process contributes to the patchy reionization foreground \citep{hu2000reionization,mukherjee2019patchy}. While efforts such as multi-wavelength observations and lensing potential reconstruction \citep{carron2017internal,krachmalnicoff2018s} are in progress to correct for the galactic and lensing foregrounds, relatively less attention is devoted to tackling the reionization foreground. And this neglect is primarily because of our lack in understanding the exact process through which the Universe reionized.

The ionized regions formed during the epoch of reionization around the ionizing sources lead to a picture where reionization is highly inhomogeneous or ``patchy''. The inhomogeneous scattering of local CMB quadrupole off these patchy free electron population (ionized regions) leads to   secondary $B$-mode polarization \citep{hu2000reionization,baumann2003small,dvorkin2009b,mortonson2010observational,su2011improved,namikawa2018constraints,mukherjee2019patchy,roy2021revised}. \cite{mortonson2010observational} found that ionized regions of size $\sim 80 \textrm{Mpc}$ could generate $B$-mode signal from reionization of comparable amplitude to lensed $B$-mode signals. More recently, \cite{mukherjee2019patchy}, \cite{paul2020} and \cite{roy2021revised} have found that the amplitude of the secondary $B$-mode from reionization depends not only on the reionization history of the Universe but also on the morphology of ionized regions. It is also well known that the morphology of these regions depend on the nature of the ionizing galaxies. For a given reionization history, the patchiness in the ionized regions increases with an increase in the minimum mass of haloes $M_{\mathrm{min}}$ which can host ionizing sources and hence, lead to a higher secondary $B$-mode signal. This implies that reliable knowledge of physical properties of the ionizing sources could play an important role in modelling the $B$-mode foregrounds, enabling a robust measurement of $r$. It is, therefore, worth carrying out a systematic study to check if reionization histories, as allowed by the available data, can be a significant obstacle in detecting the primordial gravitational waves.

In a recent effort, \cite{10.1093/mnrasl/slaa185} constrained a physical model of reionization using \emph{only} available CMB observables, i.e., the measurements of Thomson scattering optical depth $\tau$ from Planck \citep{PlanckCollaboration2018} and the kinetic Sunyaev-Zeldovich (kSZ) temperature anisotropy measurement from SPT \citep{Reichardt2020}. This allowed them to gain insights into the patchiness in the distribution of ionized regions by constraining the properties of ionizing sources. They placed an upper bound on the $B$-mode power by patchy reionization at $D^{BB,\mathrm{reion}}_{\ell=200}<18 ~\mathrm{nK}^2$ which is of the same order as the primordial contribution $D^{BB,\mathrm{prim}}_{\ell=200} \sim 5~\mathrm{nK}^2$, assuming a tensor-to-scalar ratio $r=5 \times 10^{-4}$ and tensor perturbation spectral index $n_t=0$. Both of these are orders of magnitude lower than the lensed contribution to power $D^{BB,\mathrm{lens}}_{\ell=200} \sim 10^4~\mathrm{nK}^2$. Additionally, from the constraints on parameters of reionization models, they observed that reionization cannot be driven by extremely rare sources and provided stringent constraints on the duration of reionization at $\Delta z=1.30^{+0.19}_{-0.60}$. In light of these insights about the process of reionization and lack of understanding of how significant a foreground reionization is to our observations of primordial $B$-mode signal, it becomes a natural choice to extend the works of \cite{10.1093/mnrasl/slaa185} to assess if models of reionization allowed by the current CMB observations of $\tau$ and kSZ signal can bias our inferences on $r$ for the upcoming $B$-mode observing observatories. 

The principal target of this work is therefore twofold: The first is to develop a framework to compute reionization observables and CMB anisotropies self-consistently. This would require coupling a model of reionization to codes used for computing the CMB angular power spectra. This has been carried out in the context of the semi-analytical models earlier \citep{mitra_cmb_2011,mitra2012joint,mitra2015cosmic,mitra2018cosmic,ATRI_CMB}, here we extend the method to semi-numerical models of patchy reionization. The second aim is to estimate the bias $\Delta r$ on the measurement of $r$ by future CMB experiments arising from an unaccounted-for (or incorrectly modelled) patchy reionization.

Both of these goals can be achieved by a physically motivated model of patchy reionization which can track the evolution of the intergalactic medium (IGM) in a large simulation volume. Furthermore, as we aim to explore the space of unknown parameters in as many details as possible, the model should be numerically efficient. Following our earlier works \citep{mukherjee2019patchy, paul2020, 10.1093/mnrasl/slaa185}, we simulate the patchy reionization using an efficient explicitly photon-conserving semi-numerical model of reionization, namely, Semi-numerical Code for ReIonization with PhoTon-conservation \citep[\texttt{SCRIPT};][]{choudhury2018photon}. The advantage of this model is that it allows sufficient flexibility to choose the parameters characterizing the ionizing sources and producing the CMB signals at scales of our interest. We couple the model with an existing Boltzmann solver code for CMB anisotropies, \texttt{CAMB}, and use it to make Bayesian inference on $r$ using a combination of different CMB observables, namely, the simulated $B$-mode, $\tau$, and kSZ measurements corresponding to available and future experiments. The bias $\Delta r$ is then estimated by comparing the inferred $r$ for a model of $B$-mode power spectrum which accounts for reionization correctly with a model which neglects the patchy reionization contribution to $B$-mode. The value of $\Delta r$ thus obtained would provide indications of how crucial the modelling of reionization is going to be for robust measurement of the value of $r$.

The paper is organized as follows: In Section \ref{sec:model} we briefly describe the CMB probes of reionization and the simulation of reionization with which we evaluate these probes. We confront our models with existing and upcoming probes of $\tau$ and kSZ signal and obtain constraints on the reionization history, the results of which are presented in \ref{sec:ksztauparam}. In Section \ref{sec:r_intro} we present our framework to compute the $B$-mode angular power spectrum while consistently evaluating power from primary and secondary anisotropy routines. We lay out the parameter estimation scheme to estimate $\Delta r$ and present the forecast on the bias of parameter $r$ for upcoming CMB experiments. We also discuss the implications of our findings, with focus on Stage 4 CMB missions, namely, CMB-S4 and PICO. Finally, we summarize our conclusions in Section \ref{sec:conclude}. Throughout the study we have fixed the cosmological parameters to $[\Omega_m, \Omega_b, h, n_s, \sigma_8] = [0.308, 0.0482, 0.678, 0.961, 0.829]$ \citep{PlanckCollaboration2014} which is consistent with \cite{PlanckCollaboration2018}.
\section{Secondary CMB anisotropies from patchy reionization}

\label{sec:model}

\subsection{CMB-reionization connection}
\label{subsec:cmbreion}

The process of reionization increases the free electron content of the Universe. The CMB photons Thomson scatter off these free-electron populations impacting the temperature and polarization power spectra. The three main measurable features of this impact are the Thomson scattering optical depth $\tau$, the patchy kSZ signal, and the patchy $B$-mode polarization signal \citep{hu2000reionization}. Below we briefly discuss these observables in the rest of this section.

\begin{enumerate}

\item The Thomson scattering of large-scale CMB temperature quadrupole by free electrons during the epoch of reionization creates a pronounced peak in the CMB polarization power spectra at $\ell<10$. The amplitude of this peak is quantified in terms of optical depth $\tau$ and can be defined as
\begin{equation}
    \label{eq:tau}
    \tau(z) = \sigma_T \bar{n}_H c\int^z_0 \frac{dz^\prime}{H(z^\prime)}(1+z^\prime)^2\bar{x}_e(z^\prime);~~~~
    \tau = \tau(z_{\mathrm{LSS}}).
\end{equation}
Here, $\bar{n}_H$ is the mean comoving number density of hydrogen, $\sigma_T$ is the Thomson scattering cross-section and $\bar{x}_e(z)$ denotes the global free electron fraction at a redshift $z$ with $x_e(\mbf{x},z) \equiv n_e(\mbf{x},z)/\bar{n}_H$. The redshift of the last scattering surface is denoted by $z_{\mathrm{LSS}}$.

\item The patchy kSZ signal is sourced from the Doppler shift imparted to CMB photons when they scatter off ionized bubbles with a net bulk velocity. Thus the kSZ signal is a temperature anisotropy expected to be dominant at scales corresponding to the size of ionized bubbles \citep{mcquinn2005kinetic,iliev2006ksz,Park_2013,paul2020}. Under Limber approximation \citep{Limber_1953}, accurate for the multipoles $\ell$ we are interested in, the angular power spectra of kSZ signal sourced from patchy reionization era are given as \citep{ma2002nonlinear,Park_2013,alvarez2016kinetic} 
\bear
C_\ell^{\mathrm{kSZ,reion}} &= \left(\sigma_T \bar{n}_{H} T_0\right)^2 \int \f{c~\de z^\prime}{H(z^\prime)}~\f{(1+z^\prime)^4}{\chi^2(z^\prime)} \times
\nline
& \times \e^{-2 \tau(z^\prime)}~\f{P_{q_\perp}(k = l/\chi(z^\prime), z^\prime)}{2}.
\label{eq:ksz_lim}
\ear
Here, $P_{q_\perp}$ is the power spectrum of transverse component of the Fourier transform of the momentum field $\mbf{q}(\mbf{k},z)$. The dimensionless momentum field is defined as $\mathbf{q}(\mathbf{x}, z) \equiv x_{e}(\mathbf{x}, z) \mathbf{v}(\mathbf{x}, z) / c$ where $\mathbf{v}$ is the bulk velocity field. The transverse component of momentum power spectra $P_{q_\perp}(k)$ at a wave number $k$ receives a contribution from density and velocity auto/cross power spectra at various wave modes. When simulating $P_{q_\perp}$ for finite size boxes, one misses out on the contribution of wavemodes with wavelengths larger than the size of the box. When integrated along the line of sight, missing wave modes will result in a smaller kSZ estimate when simulating kSZ for smaller box sizes \citep{iliev2007kinetic,Park_2013}. To estimate and correct for the missing power one needs information on large-scale electron density and velocity fields. In this work we include the analytical calculation of the missing power in $P_{q_\perp}$ for finite box size based on the formulation by \cite{Park_2013}. Therefore the total power in the transverse component of the momentum field is the sum of the power from the simulation box and the missing power from the analytical estimate.  For a more elaborate discussion and exact formulation for missing power used in this work refer to Appendix \ref{app:mispower}. The contribution from patchiness in the electron density during the epoch of reionization depends on the history of reionization, morphology of the ionised bubble and its spatial distribution. All these aspects impact the strength of the power spectrum of the kSZ signal induced during reionization and one needs to capture it from physics-driven reionization models. Several efforts are also made towards such studies \citep{Park_2013,paul2020,Gorce:2022cvb,Trac:2021qbn, Chen:2022lhr}. The total observed kSZ is an integrated effect which requires the need to account for the kSZ contribution from both reionization ($C_\ell^{\mathrm{kSZ, reion}}$) and post-reionization ($C_\ell^{\mathrm{kSZ, post  reion}}$) epoch. We account for the post-reionization kSZ using the scaling relations presented in \cite{shaw2012deconstructing}. Therefore the total kSZ power spectrum is evaluated as $C_\ell^{\mathrm{kSZ, tot}}=C_\ell^{\mathrm{kSZ, reion}}+ C_\ell^{\mathrm{kSZ, postreion}}$.

\item Finally, the patchy $B$-mode polarization arises as a result of Thomson scattering of  CMB temperature quadrupole off the inhomogeneous ionized field \citep{hu2000reionization}. Under Limber approximation (valid at $\ell\gtrsim 30$), the $B$-mode angular power spectra is given by \citep{hu2000reionization, Dvorkin:2008tf} 
\bear
\label{eq:BB_lim}
C_\ell^{BB,\mathrm{reion}} &= \frac{6 \bar{n}^2_{H}\sigma^2_T}{100} \int \f{c~\de z^\prime}{H(z^\prime)}~\f{(1+z^\prime)^4}{\chi^2(z^\prime)} \times
\nline
& \times \e^{-2 \tau(z^\prime)}~
P_{ee}(k = l/\chi(z^\prime), z^\prime) \frac{Q_{\mathrm{RMS}}^2}{ {2}}.
\ear
Here, $P_{ee}$ is the power spectrum of fluctuations in free electron fraction $x_e$. $Q_{\mathrm{RMS}}$ is the r.m.s of the primary quadrupole and is assumed to be constant at a value of $22~\mu K$ over the redshifts corresponding to the epoch of reionization \citep{Dvorkin:2008tf}. A computational challenge arises when simulating $B$-mode power for large angular modes. Under Limber approximation, low-$\ell$ or large angular modes will arise from wave modes of $P_{ee}(k)$ which may be of greater wavelength than the size of the box. In \cite{mukherjee2019patchy} it was noted that the ionization fluctuations at scales larger than the bubble sizes are determined by fluctuations in the density field of the haloes which produce ionizing photons. The large-scale electron density power spectrum is then a scaled form of dark matter density power spectrum with the scaling term dependent on the halo bias parameter.

\end{enumerate}

As mentioned above, we use Limber approximation to compute the angular power spectra. 
For a detailed discussion on the applicability of the approximation to our study, refer to Appendix \ref{app:limber} where we compare the Limber-approximated $B$-mode and kSZ signals to their exact counterparts.

\subsection{Reionization simulations using \texttt{SCRIPT}}

As discussed earlier, simulating and constraining parameter space of reionization using observables requires computationally efficient schemes. For Bayesian methods, physically motivated semi-numerical models are a natural choice as they are numerically efficient in parameter space exploration while allowing us to track relevant astrophysical parameters like Thomson scattering optical depth $\tau$, free electron fraction, and so on.

Considering the above arguments, we employ a semi-numerical scheme \texttt{SCRIPT} to generate ionization maps for this study. \texttt{SCRIPT} (Semi-numerical Code for ReIonization with PhoTon-conservation) is an explicitly photon conserving code for simulating the epoch of reionization \citep{choudhury2018photon}. In addition to its efficiency for Bayesian studies, the key advantage of \texttt{SCRIPT} is that the large-scale power spectrum of ionization fields converges across map resolutions. This is useful if one aims to efficiently study the large-scale properties in a simulation box. 

As a first step in simulating the CMB signals arising from reionization, we need to simulate the ionization maps given the underlying dark matter density field at redshifts of reionization. The only input parameters needed for this step are the cosmological parameters. Since we do not vary these parameters in this work, this step is needed to be carried out only once. We generate dark matter snapshots for redshifts $5 \leq z \leq 20$ by employing the 2LPT prescription in MUSIC \citep{hahn2011multi} for  box length of $512 ~ h^{-1}$ Mpc with $512^3$ particles. The collapsed mass fraction in haloes is computed using a subgrid prescription based on the conditional ellipsoidal collapse model \citep{sheth2002excursion}, see \cite{choudhury2018photon} for more details of the method.

 To obtain an ionization map at a redshift, \texttt{SCRIPT} requires two input parameters, ionizing efficiency $\zeta$ of star-forming haloes and minimum mass $M_{\mathrm{min}}$ of haloes that can host these sources. The output from \texttt{SCRIPT} is the map of ionized hydrogen fraction $x_{\mathrm{HII}}(\mbf{x},z)$.  For this study, our parameter of interest is the free electron fraction
\be
x_e(\mbf{x},z) = \chi_{\mathrm{He}}~x_{\mathrm{HII}}(\mbf{x},z)~\Delta(\mbf{x},z),
\ee
where, $\chi_{\mathrm{He}}$ is the correction factor to account for free electrons from ionized Helium and $\Delta(\mbf{x},z)$ corresponds to the dark matter overdensity. In our analysis, we consider $\chi_{\mathrm{He}}=1.08$ for $z>3$ corresponding to contribution from singly-ionized Helium and $\chi_{\mathrm{He}}=1.16$ for $z<3$ to account for free electron contribution from doubly ionized Helium. To enable us to capture the small-scale inhomogeneities, ionization maps using \texttt{SCRIPT} are generated with the best possible resolution of $2 ~ h^{-1}$ Mpc.

 In this study, we operate with the four-parameter physical model of reionization introduced in \cite{10.1093/mnrasl/slaa185}. For completeness we briefly describe the model here:  A power-law dependence of $\zeta$ and $M_{\mathrm{min}}$ on redshift is assumed to cover up for the lack of knowledge about the properties of ionizing sources at redshifts corresponding to reionization era. The parameterization for $\zeta$ and $M_{\mathrm{min}}$ is taken as following 
\be
\zeta(z) = \zeta_0 \left(\f{1 + z}{9}\right)^{\alpha_{\zeta}},~~ M_{\mathrm{min}}(z) = M_{\mathrm{min},0} \left(\f{1 + z}{9}\right)^{\alpha_{M}},
\ee
Here, $M_{\mathrm{min},0}$ is the minimum mass of haloes which can contribute to the ionizing process at redshift $z=8$ while $\zeta_0$ is the ionizing efficiency of these sources at $z=8$. The parameters $\alpha_M$ and $\alpha_\zeta$ correspond to indices of the power law. Therefore, the reionization process can be completely described by the four free parameters $\mathbf{\theta}\equiv[\log (\zeta_0),\log M_{\mathrm{min},0},\alpha_\zeta,\alpha_M]$. Given these four parameters, one can compute ionization maps with \texttt{SCRIPT} and derive the CMB observables of reionization i.e. optical depth $\tau$, the patchy kSZ signal and the patchy $B$-mode signal as described in Section \ref{subsec:cmbreion}.


\section{Revisiting constraints on reionization parameters using optical depth and kSZ}
\label{sec:ksztauparam}

In this section we discuss the parameter constraints obtained on reionization parameters using combinations of $\tau$ and kSZ CMB probes. 
 This will also inform us about the allowed ranges of reionization histories, the knowledge of which will play an important role in the construction of mock observational data with future telescopes (see Section \ref{sec:r_intro}).
 
\subsection{Parameter Constraints from current measurements}\label{subsec:ksztauparam}
We first discuss the observational constraints we plan to use in this part. The best measurement on $\tau$ is from Planck \citep{PlanckCollaboration2018} at $\tau=0.054$ with $\sigma^{\mathrm{obs}}_\tau=0.007$ inferred from the full Planck mission TT ($2 \leq \ell \leq 2500$), TE ($30 \leq \ell \leq 2000$) and EE ($2 \leq \ell \leq 2000$) data combined with the Planck CMB lensing signal ($8 \leq \ell \leq 400$). For the kSZ signal, the first $3\sigma$ measurement was made by SPT team \citep{Reichardt2020} at $D^{\mathrm{kSZ,obs}}_{\ell=3000}=\ell(\ell+1)C^{\mathrm{kSZ,obs}}_{\ell=3000} =3 ~\mu K^2$ with a $\sigma^{\mathrm{kSZ,obs}}_{\ell=3000}=1~\mu K^2$ using the temperature and polarization signal from the 2500 $\mathrm{deg}^2$ SPT-SZ and 500 $\mathrm{deg}^2$ SPT-pol surveys in the range $2000 \leq \ell  \leq 11,000$ (corresponding to angular scales of $1' \lesssim \theta \lesssim 5'$). Additionally, similar to the study by \cite{10.1093/mnrasl/slaa185}, we allow only those reionization histories which complete by redshift  ($z > 5$), consistent with constraints presented in \citep{McGreer2011,Kulkarni2019,Choudhury2020,Qin2020}. 

 We use the publicly available Markov chain Monte Carlo (MCMC) sampler available in the \texttt{Cobaya} framework \citep{Torrado_2021} to sample the set of free parameters $\mathbf{\theta}$ and compare the derived parameters ($\tau,D^{\mathrm{kSZ,tot}}_{\ell=3000}$) with the measurement data sets ($\tau^{\mathrm{obs}},D^{\mathrm{kSZ,obs}}_{\ell=3000}$). 

The main input to the MCMC code is the likelihood $\mathcal{L}$, which is calculated as 
\begin{equation}
    -2\log \mathcal{L}  =  \cc{\frac{\tau-\tau^{\mathrm{obs}}}{\sigma^{obs}_\tau}}^2 + \cc{\frac{D^{\mathrm{kSZ,tot}}_{\ell=3000}-D^{\mathrm{kSZ,obs}}_{\ell=3000}}{\sigma^{\mathrm{kSZ,obs}}_{\ell=3000}}}^2.
\end{equation}

 Apart from constraints on free parameters, we derive constraints on a set of derived parameters, namely, the amplitude of $B$-mode power spectra from patchy reionization $D^{BB,\mathrm{reion}}_{\ell=200}$ at a multipole $\ell = 200$, the redshifts $[z_{25},z_{50},z_{75}]$ corresponding to the mass averaged ionized fraction $Q_{\mathrm{HII}}=[0.25,0.50,0.75]$ respectively and the duration of reionization defined as $\Delta z=z_{25}-z_{75}$ from the model for each sample of free parameters.

\begin{figure*}
    \centering
    \includegraphics[width=\linewidth]{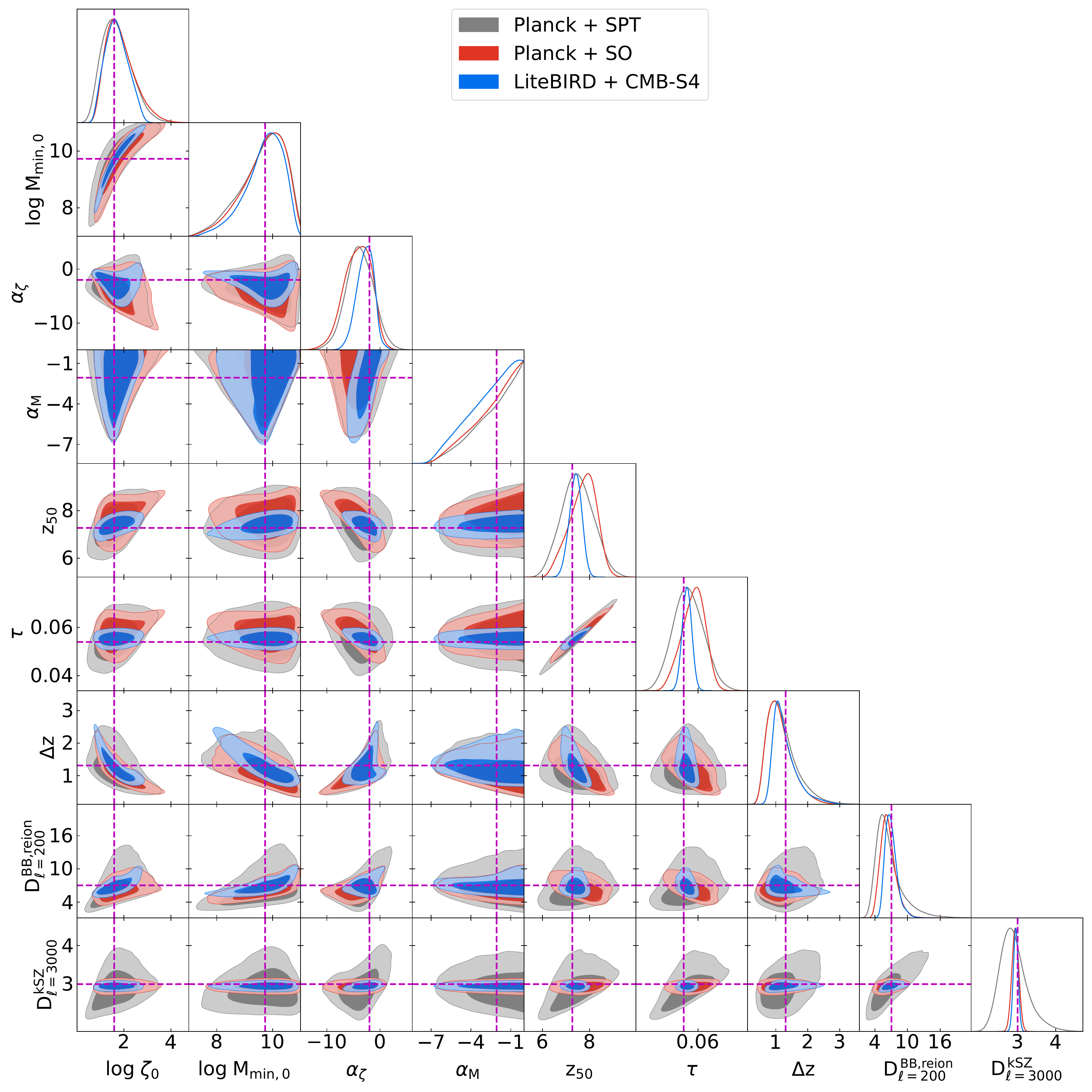}
    \caption{The posterior distribution of free and derived parameters of the reionization model for different combinations of data sets as mentioned in the figure legend has been presented. The posteriors show both $68\%$ and $95\%$ contours in the two-dimensional posterior plots. The dashed magenta lines denote the input values used for forecasting.}
    \label{fig:taukszall}
\end{figure*}

The parameter constraints obtained using Planck \citep{PlanckCollaboration2018} and SPT \citep{Reichardt2020} are shown in Table \ref{tab:tauksz_const} with one and two-dimensional posterior distribution shown in Figure \ref{fig:taukszall}. 
We find that the constraints are slightly different compared to those provided by \cite{10.1093/mnrasl/slaa185}. This is mainly because the contribution to $P_{q_{\perp}}(k)$ arising from wavemodes with wavelengths larger than the simulation box was not taken into account by \citet{10.1093/mnrasl/slaa185}. In spite of these differences, the overall conclusions remain the same. The important takeaways are: we see that the data prefer $M_{\mathrm{min,0}}\gtrsim 10^9 ~M_\odot$ (at $68\% $ C.L.) indicative of suppressed star formation in low mass haloes as a result of radiative feedback at $z\sim 8$ and that $\alpha_\zeta$ prefers a negative constraint. This could be indicative of more efficient cooling and star formation or increased escape fraction at lower redshifts. Finally, we obtain a tight constraint on the width of reionization $\Delta z=1.19^{+0.27}_{-0.53}$.

\begin{figure}
	\includegraphics[width=\columnwidth]{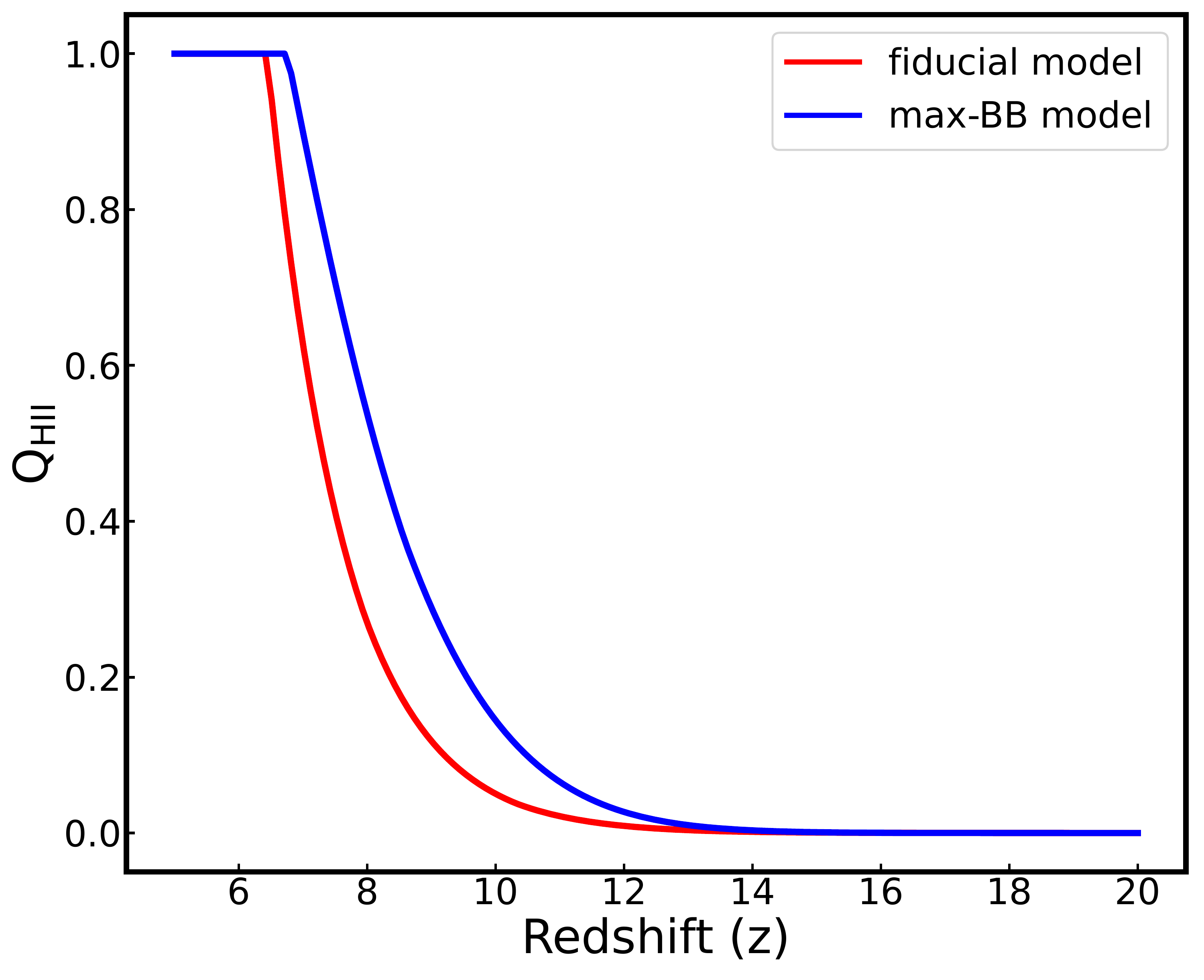}
    \caption{Redshift evolution of mass-averaged ionized fraction $Q_{\rm{HII}}(z)$ for the fiducial and max-BB models of reionization. See the text for a description of the models.}
    \label{fig:ionizationfrac}
\end{figure}

The best fit model obtained from the analysis is $\rr{\log (M_{\mathrm{min,0}})=9.73, \log (\zeta_0)=1.58, \alpha_M=-2.06, \alpha_\zeta=-2.01}$. We use this model as the fiducial model of reionization when forecasting for measurements with future probes (see Section \ref{subsec:bmoderesults}). The value of $\tau$ for this fiducial model is $0.0540$. The redshift evolution of the global mass-averaged ionization fraction $Q_{\rm{HII}}(z) \equiv \langle x_{\mathrm{HII}}(\mbf{x},z)~\Delta(\mbf{x},z) \rangle$ is shown in Figure \ref{fig:ionizationfrac} (red curve). In addition, we also use a model, named max-BB, which has the maximum $D^{BB,\mathrm{reion}}_{\ell=200}$  among models that are allowed within the $3\sigma$ confidence levels (see Section \ref{subsec:maxBB}). This model has parameters $\rr{\log (M_{\mathrm{min},0})=10.39, \log (\zeta_0)=2.48, \alpha_M=-0.76, \alpha_\zeta=3.58}$. The value of $\tau$ in this case is $0.0627$, the corresponding $Q_{\rm{HII}}(z)$ is shown as the blue curve in Figure \ref{fig:ionizationfrac}. We find that the max-BB model has an early but more gradual evolution of ionization fraction compared to the fiducial model.

\begin{table}
    \centering
    \caption{Constraints of reionization model parameters obtained from MCMC analysis using current observations from Planck and SPT. The first four rows correspond to the free parameters of the model while the rest of the parameters are the derived parameters. Uniform priors have been assumed for the free parameters in the prior range mentioned in the second column.}
    \begin{tabular}{|c|c|c|}
    \hline
    Data & Prior & Planck + SPT\\
    Parameter&  & $68\% $ limits \\
    \hline
    $\log (\zeta_0)$ & $[0,\infty]$  &   $1.70^{+0.49}_{-0.76}$  \\[0.075cm]
    $\log (M_{\mathrm{min,0}})$& $[7.0,11.0]$ & $9.65^{+1.02}_{-0.49}$         \\[0.1cm]
    $\alpha_\zeta$  & $[-\infty,\infty]$   & $-3.81^{+2.58}_{-2.52}$ \\[0.1cm]
    $\alpha_M$ & $[-\infty,0]$   & $>-2.78$       \\[0.1cm]
    \hline
    Derived Parameters &&\\
    $\tau$&   & $0.0559^{+0.0062}_{-0.0067}$\\[0.1cm]
    $z_{\mathrm{25}}$&& \pc{8.24}{0.66}{0.66} \\[0.1cm]
    $z_{\mathrm{50}}$&& \pc{7.49}{0.69}{0.68}\\[0.1cm]
    $z_{\mathrm{75}}$&& \pc{7.04}{0.76}{0.46}\\[0.1cm]
    $\Delta z$ &   & $1.19^{+0.27}_{-0.53}$      \\[0.1cm]
    $D^{\mathrm{kSZ}}_{l=3000}$($\mu K^2$)&  & $2.90^{+0.26}_{-0.41}$ \\[0.1cm]
    $D^{BB,\mathrm{reion}}_{\ell=200}$($n K^2$)  & & $6.60^{+1.13}_{-2.73}$ \\
    \hline
    \end{tabular}
    \label{tab:tauksz_const}
\end{table}

\subsection{Forecasts for future CMB experiments aiming to measure kSZ effect}
\label{sec:taukszforecast}
\begin{table}
    \centering
    \caption{Noise specification for the upcoming ground-based CMB experiments for their LAT configuration}
    \begin{tabular}{c|c|c|c|c}
    \hline
    Mission & Frequency & $\Delta T$& Beam & $f_{sky}$  \\
    &(GHz)& ($\mu \mathrm{K}-\rm arcmin$) & ($\rm arcmin$) &\\
    \hline
    SO LAT(goal) & 145 & 6.3 & 1.4 & 0.4    \\ [0.1cm]
    CMB-S4 & 150 & 1.8 & 1.0 & 0.7\\[0.1cm]
    \hline
    \end{tabular}
    \label{tab:noise_spec_LAT}
\end{table}

We extend our analysis of constraints on reionization parameters to future CMB measurements which aim to make high fidelity temperature and polarization power spectra and measure the observables, kSZ signal at $\ell=3000$ and optical depth $\tau$. Future experiments LiteBIRD \citep{suzuki2018litebird} and PICO \citep{hanany2019pico} aim to measure the reionization bump ($\ell<10$) of the $E$-mode CMB polarization to constraint $\tau$ with $\sigma^{\mathrm{obs}}_\tau=0.002$. In the future, we also expect to make sensitive measurements of kSZ with Simons Observatory (SO) \citep{Ade_2019} and CMB-S4 \citep{Abazajian2019}. The $\sigma^{\mathrm{kSZ}}_\ell$ used to forecast constraints in our analysis for upcoming experiments is calculated using LAT configuration noise specification provided in Table \ref{tab:noise_spec_LAT} along with other sources of noise as specified for Equation (4) of \cite{10.1093/mnrasl/slaa185}.

Considering the mission timelines we propose the following scenarios of probes for measurements of $\tau$ and kSZ to forecast constraints on the reionization model parameter space: 
\begin{itemize}
\item \textbf{Planck + SO:} $\tau$ measurement using Planck with projected kSZ measurement with the upcoming SO,
\item \textbf{LiteBIRD + CMB-S4:} projected $\tau$ measurement with LiteBIRD with kSZ using CMB-S4.
\end{itemize}
We use the fiducial reionization model of the previous section to create mock observations of the upcoming facilities.

 Comparative posteriors of free and derived parameters for different combinations of $\tau$ and kSZ measuring experiments are shown in Figure \ref{fig:taukszall}. The parameter constraints corresponding to the reionization model can be seen in Table \ref{tab:taukszforecast}. It is obvious from the figure that the parameter space gets more constrained with the upcoming more sensitive experiments. In Table \ref{tab:taukszforecast} we see that, with improved kSZ observations from SO, the error bar on $\tau$ reduces marginally to $\sim 0.005$ from $\sim 0.0065$. While with LiteBIRD measurement of $\tau$ and kSZ measurement of CMB-S4, the error bars on $\tau$ would reduce to $\sim 0.002$ while that on $D^{\mathrm{kSZ}}_{\ell=3000}$ is $\sim 0.06$. Tight constraints for $\tau$ and $D^{\mathrm{kSZ}}_{\ell=3000}$ translate to tighter constraints on the properties of the ionizing sources. This further helps us to gain insight into spatial inhomogeneities and the evolution of global properties of the reionization era. For reference, with each iteration of future experiments, we will be able to achieve tighter constraints on global ionization history through $z_{25},z_{50},z_{75}$
and patchiness in electron fraction field along the line-of-sight (LOS) through constraints on $D^{BB,\mathrm{reion}}_{\ell=200}$.

\begin{table}
    \centering
    \caption{Forecasts on reionization model and derived parameters for the upcoming CMB experiments. The first four rows correspond to the free parameters of the model while the rest of the parameters are the derived parameters. The free parameters are assumed to have the same priors as Table \ref{tab:tauksz_const}. The second column shows the input value used to construct the mock data based on which forecasts are made}    
    \begin{tabular}{|c|c|c|c|}
    \hline
    Data & Input & Planck + SO & LiteBIRD + CMB-S4\\
    Parameter& Model & $68\% $ limits & $68\% $ limits \\
    \hline
    $\log \zeta_0$  &1.58  & \pc{1.85}{0.41}{0.74} & \pc{1.69}{0.42}{0.55}  \\[0.1cm]
    $\log \cc{M_{min,0}}$& 9.73 & \pc{9.68}{0.96}{0.47} & \pc{9.69}{0.84}{0.43} \\[0.1cm]
    $\alpha_\zeta$ & $-2.01$ & \pc{-4.24}{2.91}{2.24} & \pc{-2.79}{1.79}{1.39}\\[0.1cm]
    $\alpha_M$ & $-2.06$ & $>-2.83 $& $>-3.19$ \\[0.1cm]
    \hline
    Derived Parameters &&&\\
    $\tau$ &$0.0540$&\pc{0.0579}{0.0055}{0.0043}& \pc{0.0553}{0.0020}{0.0020} \\[0.1cm]
    $z_{\mathrm{25}}$ & $8.09$ &\pc{8.43}{0.47}{0.39}& \pc{8.22}{0.18}{0.21}\\[0.1cm]
    $z_{\mathrm{50}}$ & $7.27$ &\pc{7.71}{0.66}{0.47}&\pc{7.40}{0.28}{0.25}\\[0.1cm]
    $z_{\mathrm{75}}$ & $6.78$ &\pc{7.30}{0.81}{0.55}& \pc{6.93}{0.42}{0.29} \\[0.1cm]
    $\Delta z$ & $1.31$ &\pc{1.13}{0.25}{0.47}& \pc{1.29}{0.18}{0.44} \\[0.1cm]
    $D^{\mathrm{kSZ}}_{\ell=3000}$($\mu K^2$) & $3.00$ &\pc{2.94}{0.09}{0.09}& \pc{2.95}{0.06}{0.07} \\[0.1cm]
    $D^{BB,\mathrm{reion}}_{\ell=200}$($n K^2$) & $7.03$&\pc{6.44}{1.03}{1.57}& \pc{6.99}{0.78}{1.27}\\[0.1cm]    
    \hline
    \end{tabular}

    \label{tab:taukszforecast}
\end{table}

\section{Inferring tensor to scalar ratio including reionization foregrounds}\label{sec:r_intro}

Until now, we have been considering measurements of $\tau$ and $D_{\ell}^{\mathrm{kSZ}}$. We now get to the main aim of this work, i.e., to investigate the effects of patchy reionization on the detection of the $B$-modes from primordial gravitational waves. Neglecting the contribution of $B$-mode power generated during patchy reionization will lead to a bias in the mean value of the inferred tensor to scalar ratio $r$. For the upcoming CMB experiments, it is thus essential to study if the bias is sufficient to mislead the estimation of $r$.

\subsection{Simulating the $B$-mode angular spectra from CMB}
\label{sec:bmode_eval}

\begin{figure}
	\subfloat[]{\includegraphics[width=\columnwidth]{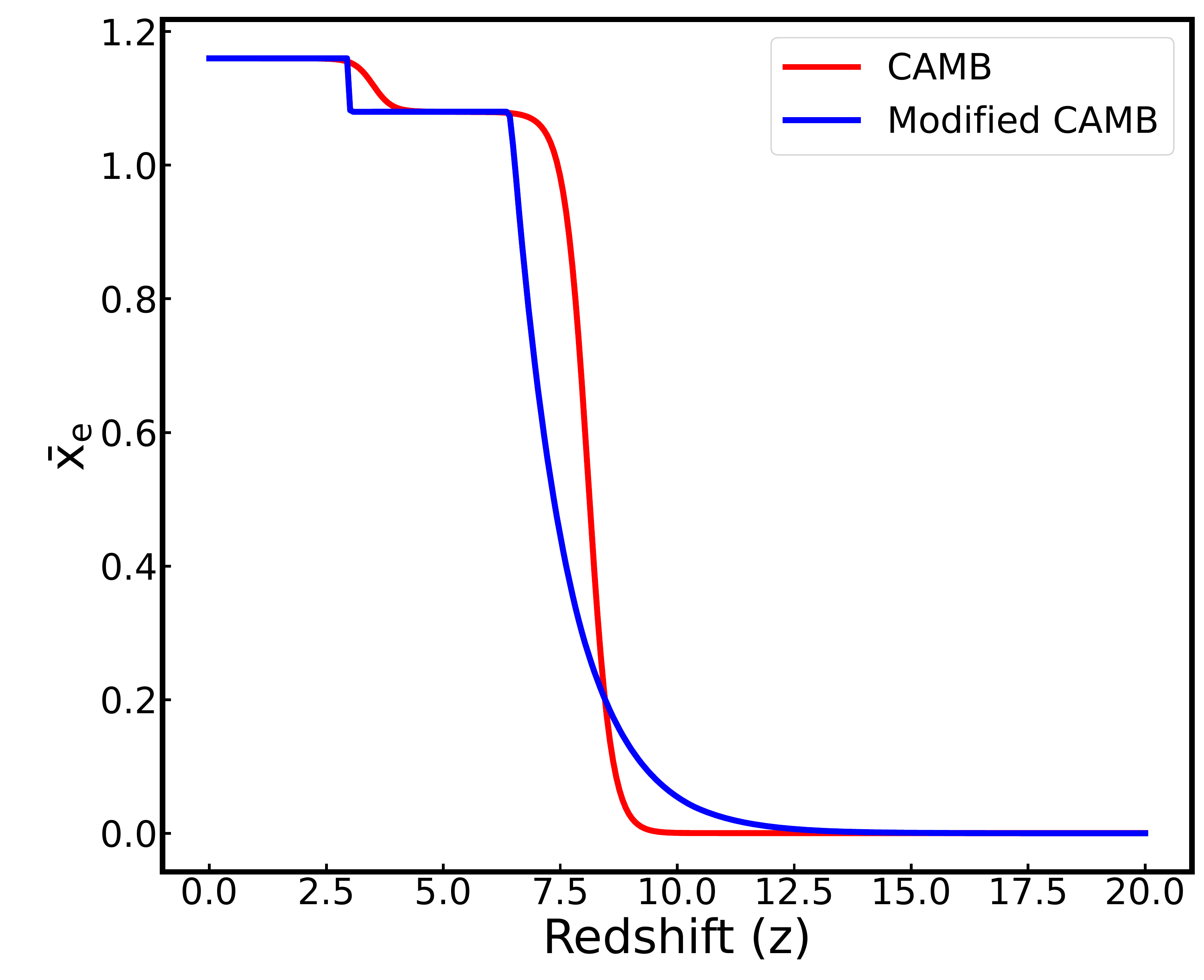}\label{cambmod}}\\
	\subfloat[]{\includegraphics[width=\columnwidth]{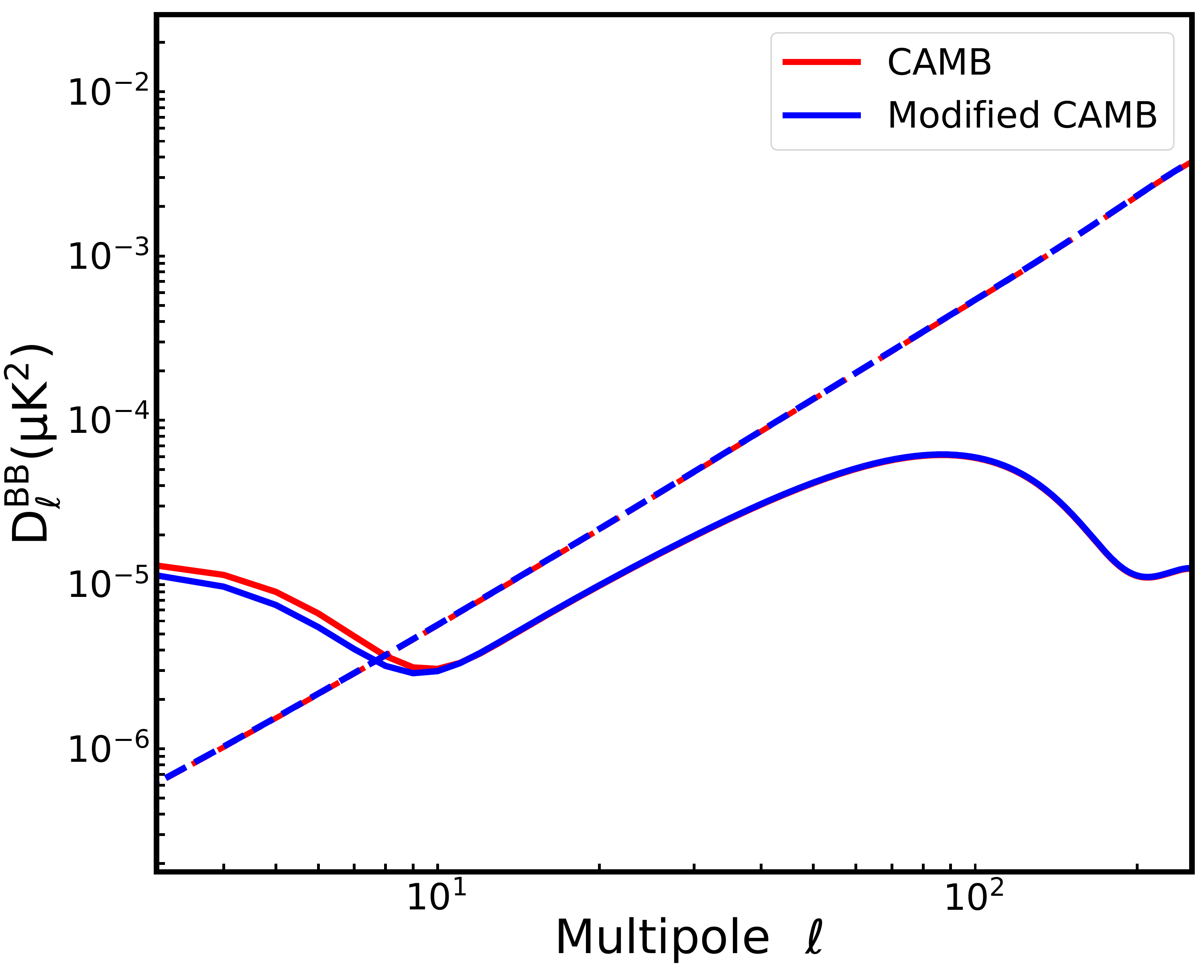}\label{cambmodxe}}
    \caption{Top panel: Comparison of mass-averaged free electron fraction, $\bar{x}_e$ (top panel) and $B$-mode power spectra (bottom panel) obtained from CAMB (in red) and modified routine of CAMB (in blue) has been presented. The default CAMB routine employs a $\tanh$ model of reionization corresponding to the input value of $\tau=0.054$ consistent with the optical depth obtained for our fiducial model. For the modified CAMB routine the fiducial model of reionization is taken as input. Bottom Panel: we show the primordial $B$-mode power spectra as solid lines and the lensed modes as dashed lines. The $B$-mode power is generated for $r=5\times 10^{-4}$ and $A_{\mathrm{lens}}=0.15$.}
    \label{fig:cambmod}
\end{figure}

We first present a self-consistent framework to compute the $B$-mode angular power spectrum for arbitrary reionization histories. The $B$-mode signal has three major contributors, the primordial gravitational wave ($C^{BB,\textrm{prim}}_{\ell}$),  the lensed scalar modes ($C^{BB,\textrm{lens}}_{\ell}$) and finally the $B$-mode arising from the patchiness of ionized fields in the era of reionization ($C^{BB,\textrm{reion}}_{\ell}$). The total $B$-mode  power observed by us is given as
\begin{equation}
\label{eq:mockBmode}
    C^{BB}_{\ell}=C^{BB,\textrm{prim}}_{\ell}+  A_{\textrm{lens}}~C^{BB,\textrm{lens}}_{\ell} + C^{BB,\textrm{reion}}_{\ell},
\end{equation}
where $A_{\textrm{lens}}$ is the residual lensing amplitude after delensing the signal. Let us discuss modeling each of these components one by one:

\begin{itemize}

\item The patchy reionization component $C^{BB,\textrm{reion}}_{\ell}$ is computed using \texttt{SCRIPT} which provides the ionized field power spectrum $P_{ee}(k, z)$. This serves as an input to Equation (\ref{eq:BB_lim}) for calculating the angular power spectrum. The computation of this component requires knowledge of the free parameters $\zeta_0, M_{\mathrm{min},0}, \alpha_{\zeta}, \alpha_M$.

\item The primary signal $C^{BB,\textrm{prim}}_{\ell}$ is computed using a Boltzmann solver \texttt{CAMB} \citep{lewis2000efficient,howlett2012cmb}. The computation of this component requires the values of parameters related to the primordial tensor power spectrum $r$ and $n_t$ and also $\tau$. While $r$ and $n_t$ are input parameters, the value of $\tau$ depends on the reionization history implied by the parameters related to reionization. 

In the default version of \texttt{CAMB}, a user could either specify the mid-point of reionization $z_{\mathrm{rei}}$ or the Thomson scattering optical depth $\tau$ to tune the default $\tanh$ ionization history. To ensure consistencies between the primordial and patchy reionization signals, we modify \texttt{CAMB} to account for arbitrary reionization histories. In this Modified CAMB, we use the ionization history of IGM evaluated as per the physical model of reionization with \texttt{SCRIPT}. In particular, we modify the code \texttt{reionization.f90} which describes the reionization module in \texttt{CAMB}. Additionally, in consistency with \texttt{SCRIPT} to account excess of electrons due to Helium ionization we use, $\chi_{\mathrm{He}}=1.16$ for $z<3$ and $\chi_{\mathrm{He}}=1.08$ otherwise. In Figure \ref{fig:cambmod} we present a comparison of mass-averaged free electron fraction and $B$-mode power spectra from default \texttt{CAMB} and Modified \texttt{CAMB} routines. Note that the two models have the same $\tau$. Interestingly, accounting for a general reionization history implied by our fiducial model of reionization leads to a slightly different $D_{\ell}^{BB}$ at low multipoles (around the reionization ``bump'').

\item The lensing signal $C^{BB,\textrm{lens}}_{\ell}$ too is computed by the Modified \texttt{CAMB} for any general reionization history.

\end{itemize}

\begin{figure*}
    \includegraphics[width=1.2\linewidth, trim=18cm 9cm 0cm 3cm, clip,scale=0.1]{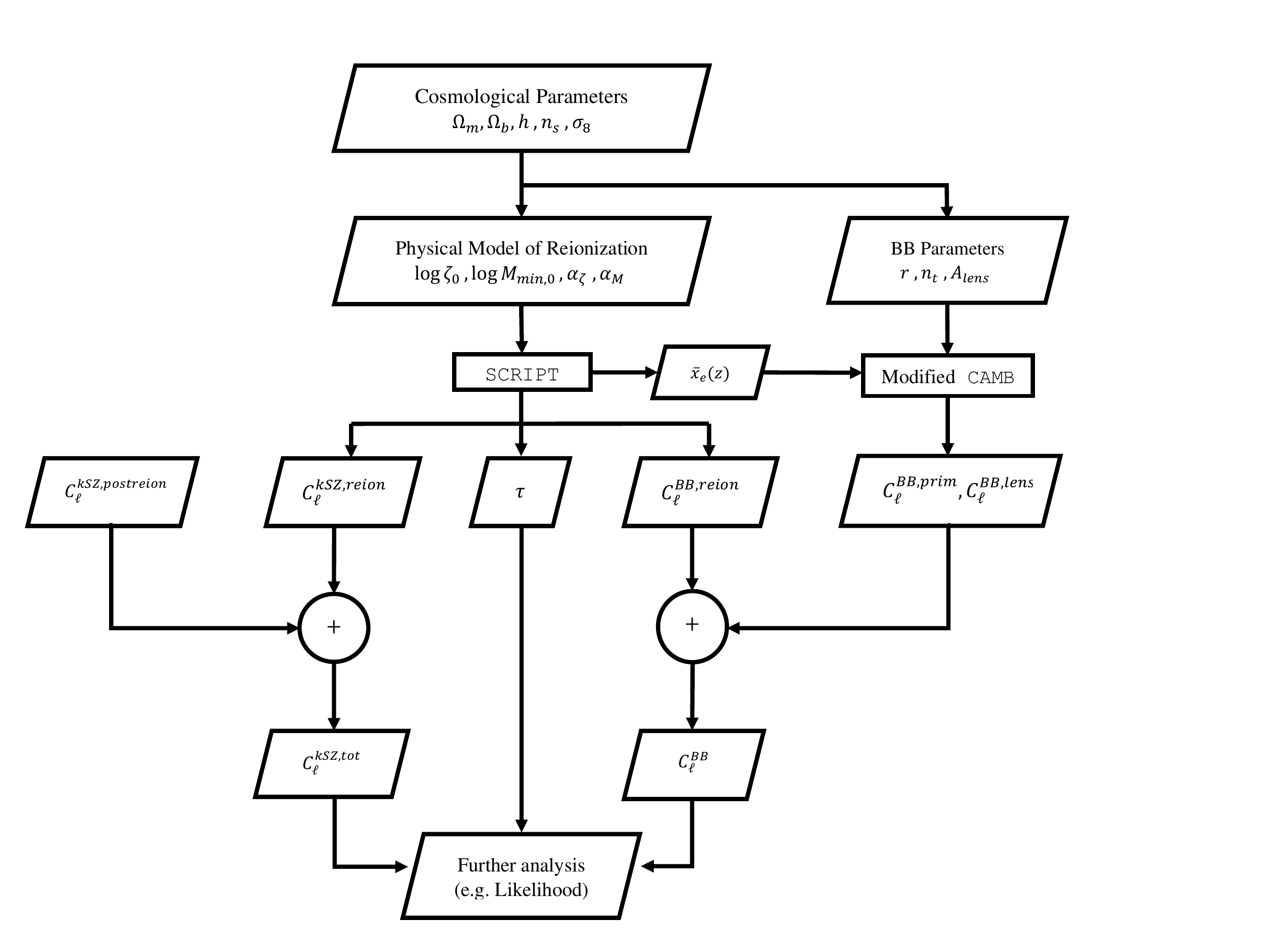}
    \caption{Flowchart to depict the framework developed in this work to compute CMB observables related to patchy reionization.}
    \label{fig:BBtotal}
\end{figure*}

A flowchart diagram summarizing the framework for computing the CMB observables for arbitrary reionization histories in a self-consistent manner is shown in Figure \ref{fig:BBtotal}. Our code requires the cosmological and reionization parameters as user inputs (along with the specifications of the simulation volume and resolution) and produces the resulting reionization history and all the relevant CMB observables. In addition, the code is capable of providing several other high-$z$ observables (e.g., galaxy luminosity function, the 21~cm maps), however, we will not discuss these in this paper.

\subsection{Likelihood and Bayesian analysis}
\label{subsec:r_likelihood}

To understand the potential impact of reionization models allowed by current CMB measurements on future CMB experiments, we make a Bayesian inference of $r$ for upcoming CMB experiments. In this effort, we use a different combination of $\tau$, kSZ, and $B$-mode power spectra probes based on the current estimates of the timeline for different missions. For each upcoming $B$-mode measuring experiment we position ourselves in time to forecast for the measurement with a choice of the best available $\tau$ and kSZ measurement. We propose four combinations for such an analysis:

\begin{itemize}
    \item Case \textbf{SO+:} Planck ($\tau$) +SO (kSZ) +SO (BB) [expected availability $\sim 2024$]
    \item Case \textbf{LiteBIRD+:} Planck ($\tau$) +SO (kSZ) + LiteBIRD (BB) [expected availability $\sim 2028$]
    \item Case \textbf{CMB-S4+:} LiteBird ($\tau$) +CMB-S4 (kSZ) + CMB-S4 (BB) [expected availability $\sim 2030$]
    \item Case \textbf{PICO+:} LiteBIRD ($\tau$) + CMB-S4 (kSZ) + PICO (BB) [expected availability sometime in the next decade]
\end{itemize}

 \begin{table}
    \centering
    \caption{Specifications of the CMB experiments aiming to target large scale $B$-modes for their SAT configuration}
    \begin{tabular}{lcccc}
   \hline
   Experiment  & $\Delta_P$  & $\Theta_{\text{FWHM}}$  & $f_{\mathrm{sky}}$& Delensing \\
    & ($\mu \text{K-arcmin}$) & ($\text{arcmin}$) & & $1 - A_{\mathrm{lens}}$\\
    \hline
    SO & 2.7 & 30.0 & 0.1 & $70\%$\\
    LiteBIRD & 2.4 & 30.0 & 1.0& $70\%$\\
    CMB-S4  & 1.5 & 30.0 & 0.7& $85\%$\\
    PICO & 0.87 & 7.9 & 1.0& $85\%$\\
    \hline
    \end{tabular}
    \label{tab:observatory_beam}
\end{table}

For the choice of a combination of probes or Cases as discussed above, we sample the parameters $\mathbf{\theta} \equiv  \left\{\log(M_{\mathrm{min}, 0}),\log(\zeta_0), \alpha_{\zeta}, \alpha_M, r \right\}$ and obtain posteriors using  MCMC sampler  in the \texttt{Cobaya} framework.
The form of likelihood we use to infer $r$ is given as:                                                                             
\begin{equation}
\begin{aligned}
    &-2\log \mathcal{L}  =  \cc{\frac{\tau-\tau^{\mathrm{obs}}}{\sigma^{obs}_\tau}}^2 + \cc{\frac{D^{\mathrm{kSZ,tot}}_{\ell=3000}-D^{\mathrm{kSZ,obs}}_{\ell=3000}}{\sigma^{\mathrm{kSZ,obs}}_{\ell=3000}}}^2+\\ &\sum^{\ell_{max}}_{\ell,\ell^\prime=\ell_{min}}\cc{\tilde{C}_\ell^{BB}-C_\ell^{BB}}\Sigma^{-1}_{\ell \ell^\prime}\cc{\tilde{C}_{
    \ell^\prime}^{BB}-C_{\ell^\prime}^{BB}}
\end{aligned}
\end{equation}

Here $\tilde{C}^{BB}_\ell$ represents the mock data power spectrum, ${C}^{BB}_\ell$ represents the model data power spectrum, and $\Sigma_{\ell\ell^\prime}$ represents the covariance matrix of the $B$-mode angular power spectrum:
\begin{equation}\label{eq:noisecov}
        \Sigma_{\ell \ell^\prime}=\frac{2}{f_{\mathrm{sky}}(2\ell+1)}\cc{\tilde{C}^{BB}_\ell+N_\ell}^2\delta_{\ell \ell^\prime}
\end{equation}
To calculate the elements of the covariance matrix, one must know the mock $B$-mode power $\tilde{C}^{BB}_\ell$ and the instrument specifications chiefly noise power spectra $N_\ell$ and the fraction of sky accessible to instrument $f_{\mathrm{sky}}$. These instrument specifications are presented in Table~\ref{tab:observatory_beam}. For Simons Observatory we have used instrumental noise corresponding to Small Aperture Telescope (SAT) configuration for the 93 GHz frequency channel \citep{Ade_2019}. For CMB-S4 we consider noise corresponding to 30 arcmin beam width for the case assuming all the detectors were concentrated at 150 GHz \citep{Abazajian2019}. For LiteBIRD we consider the total sensitivity with the angular resolution of $\sim 30$ arcmin at 150 GHz \citep{hazumi2019litebird}. Finally, for PICO, we consider the sensitivity corresponding to the final polarization combined map noise level equivalent to 3300 Planck missions \citep{hanany2019pico}.

The mock (template) of $\tilde C^{BB}_\ell$ power spectrum, see Equation \eqref{eq:mockBmode}, has contribution of $B$-mode power spectra from primordial gravitational waves $\CBB{\mathrm{prim}}$, lensing contribution $\CBB{\mathrm{lens}}$ and the $B$-mode angular power arising from the patchy reionization $\CBB{\rm reion}$.

The generation of the mock data requires one to make choices for the model parameters. We list our choices below:

\begin{enumerate}


\item We explore two values of $r$ while generating the mock $C^{BB,\mathrm{prim}}_\ell$, namely, $r = 0.001$ and $r = 5 \times 10^{-4}$. These values are much less than the present upper limits on the parameter and are typical of what the upcoming experiments aim to detect \citep{Abazajian2019,hanany2019pico}. Throughout the analysis, we assume that the spectral index of tensor perturbations, $n_t=0$. 

\item The lensing contribution requires the value of $A_{\mathrm{lens}}$, which depends on the experiment under consideration. The default values are given in Table \ref{tab:observatory_beam}. In addition to the default values, we also consider a rather optimistic case where we take $1 - A_{\mathrm{lens}} = 95\%$ \citep{diego2020comparison}. We also consider a case that is idealized with $1 - A_{\mathrm{lens}} = 100\%$. This corresponds to a situation where all the lensing signal has been subtracted and is included only to understand the effects of residual lensing signal on our conclusions. 

\item For the reionization model, our fiducial case is the best-fit model from our analysis in Section \ref{sec:ksztauparam}. In addition, we consider another model which produces the maximum $C^{BB,\mathrm{reion}}_{\ell=200}$ among those allowed within the $3 \sigma$ limits of the present constraints. We call this variant the ``max-BB'' reionization model.

\end{enumerate}

In order to understand the importance of patchy reionization while estimating the parameter $r$, we consider two cases while recovering the parameters using the mock spectra:
\begin{itemize}
    \item In the first case, we assume that the model used for comparing with the mock data does \emph{not} include the reionization contribution to $B$-mode power, i.e.,
    \begin{equation}
      \label{eq:template-reion}
         \text{Template} -  C^{BB,\mathrm{reion}}_\ell : \; C^{BB}_\ell=C^{BB,\mathrm{prim}}_\ell+A_{\rm lens}C^{BB,\mathrm{lens}}_\ell
    \end{equation}
    Since the mock data contains the contribution from the patchy reionization while the model does not, this case is expected to lead to a bias in the recovery of $r$.
    
    \item In this case, we include the patchy reionization contribution to the model
    \begin{equation}
      \label{eq:template}
         \text{Template : \;}  C^{BB}_\ell=C^{BB,\mathrm{prim}}_\ell+A_{\rm lens}C^{BB,\mathrm{lens}}_\ell+C^{BB,\mathrm{reion}}_\ell
    \end{equation}
    This case corresponds to a fair comparison between the mock data and the model and should be able to recover the input $r$ (within error bars).
\end{itemize}
The key idea here is to estimate the bias in the inferred  $r$ for the above models given as
\be
\f{\Delta r}{\sigma} \equiv \f{\cc{r_{\textrm{Template}}-r_{\textrm{Template} - C^{BB,\mathrm{reion}}_\ell}}}{\sigma_{r_{\textrm{Template}}}},
\ee
where $r_{\textrm{Template}}$ is the value of $r$ inferred with full template corresponding to Equation \eqref{eq:template}, $r_{\textrm{Template} - C^{BB,\mathrm{reion}}_\ell}$ is the one with the template without patchy reionization corresponding to Equation \eqref{eq:template-reion}  and $\sigma_{r_{\textrm{Template}}}$ is the statistical error on $r$ for the analysis with the full template. In cases of ground-based observatories like Simons Observatory (SO) and CMB-S4, the low $\ell$ modes are inaccessible, therefore, the multipole range considered in such cases is from $\ell=[52,252]$. This is not an issue for space-based missions that have access to $\ell$ modes from $\ell_{\mathrm{min}}=2$, and the multipole range considered are $\ell=[2,252]$. As was noted, to reduce the time complexity the patchy $B$-mode angular power spectrum is evaluated under Limber approximation which is valid for multipoles of $\ell \gtrsim 30$. This may raise concerns regarding our study of the bias $\Delta r$ with regard to space-based missions for which we evaluate the patchy $B$-mode for multipole as low as $\ell=2$. In Appendix \ref{app:limber} we show that the $B$-mode power evaluated through Limber approximation is underestimated at multipoles ($\ell \lesssim 30$). Therefore, we argue that any bias estimate $\Delta r$ from our study should always be treated as a lower limit in principle. We further note that a substantial difference in the bias estimate is highly unlikely as the amplitude of the primordial signal will be greater than the patchy $B$-mode signal for the concerned range of multipoles by a few orders of magnitude at multipoles $\ell \lesssim 30$.

\begin{figure}
	\includegraphics[width=\columnwidth]{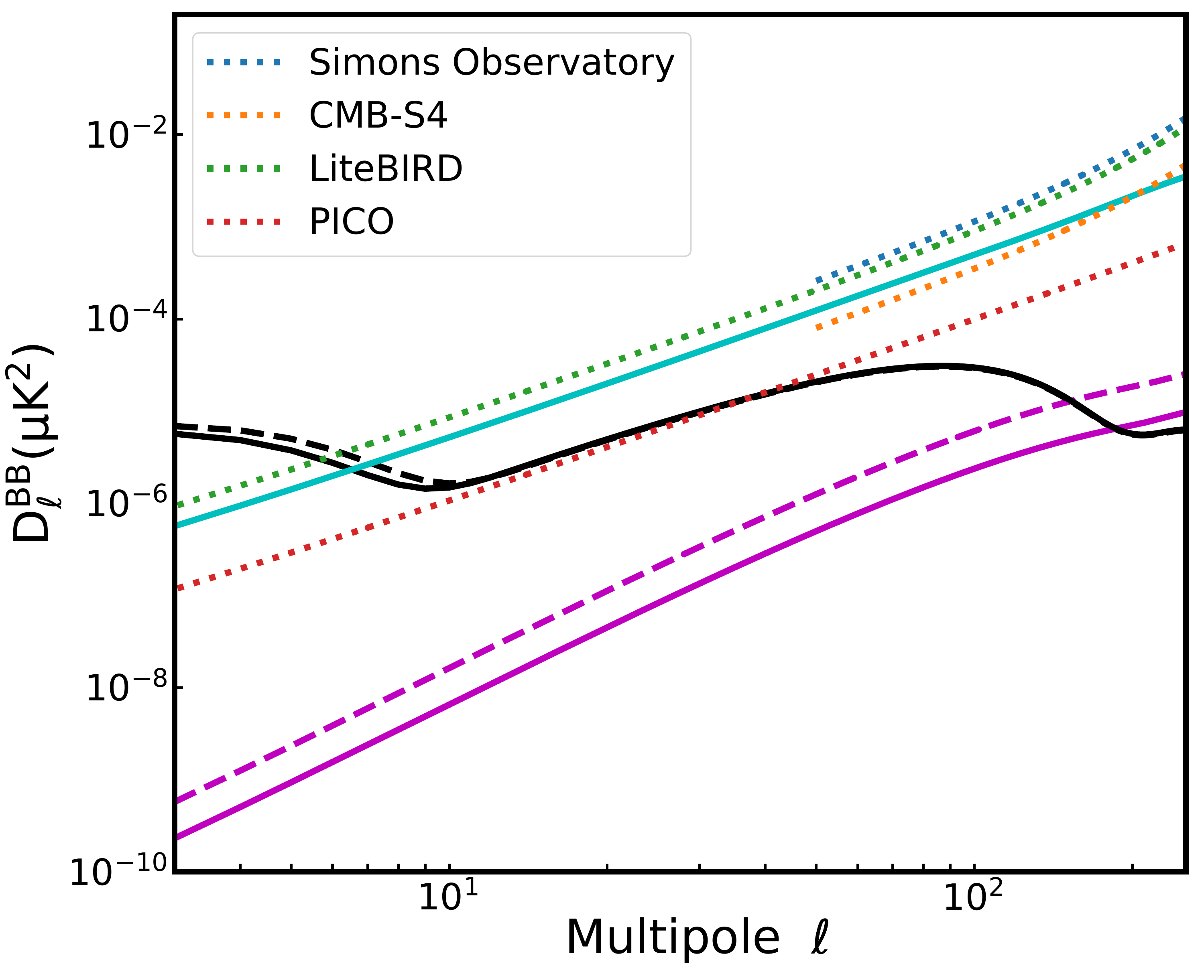}
    \caption{The angular power spectrum of $B$-mode from patchy reionization in magenta for our fiducial model (solid) and max-BB (dashed) of reionization has been shown. The corresponding primordial $B$-mode power spectra (black solid and dashed curves) for tensor to scalar ratio of $5\times 10^{-4}$ are also shown. The cyan curve denotes power due to weak lensing (with lensing amplitude of $1 - A_{\mathrm{lens}} = 0.85$). The dotted lines represent the instrumental noise power spectra for Simons Observatory (blue), CMB-S4 (orange), LiteBIRD (green), and PICO (red) corresponding to the noise specifications shown in Table~\ref{tab:observatory_beam}.}
    \label{fig:BBobs}
\end{figure}

As a reference, we show the different $B$-mode power contribution into the mock $B$-mode power spectra in Figure \ref{fig:BBobs}. The solid and dashed lines correspond to the power spectra arising from primordial $B$-mode power spectra (in black), lensed $B$-mode power spectra (in cyan), and $B$-mode power from patchiness in the reionization process (in magenta). In addition, we present noise spectra (in dotted lines)  of observatories aiming to observe $B$-mode power spectra corresponding to specifications presented in Table \ref{tab:observatory_beam}.

Before moving on to the Bayesian analysis, let us understand the scales where the reionization signal can be important. In Figure \ref{fig:BBobs} we find that the patchy $B$-mode power arising during the reionization era for our fiducial model of reionization (in a solid magenta curve) is comparable to or greater than the primordial $B$-mode power ($r=5\times 10^{-4}$) for multipoles $\ell \gtrsim 100$. While lensed $B$-modes are a dominant contribution at multipoles $\ell \gtrsim 10$, nevertheless for tensor-to-scalar ratio $r\lesssim 10^{-3}$ and with improved delensing strategies, patchy $B$-mode may appear a significant foreground. 

\subsection{Estimation of bias in $r$: fiducial reionization model}
\label{subsec:bmoderesults}

\begin{figure}
	\includegraphics[width=\columnwidth]{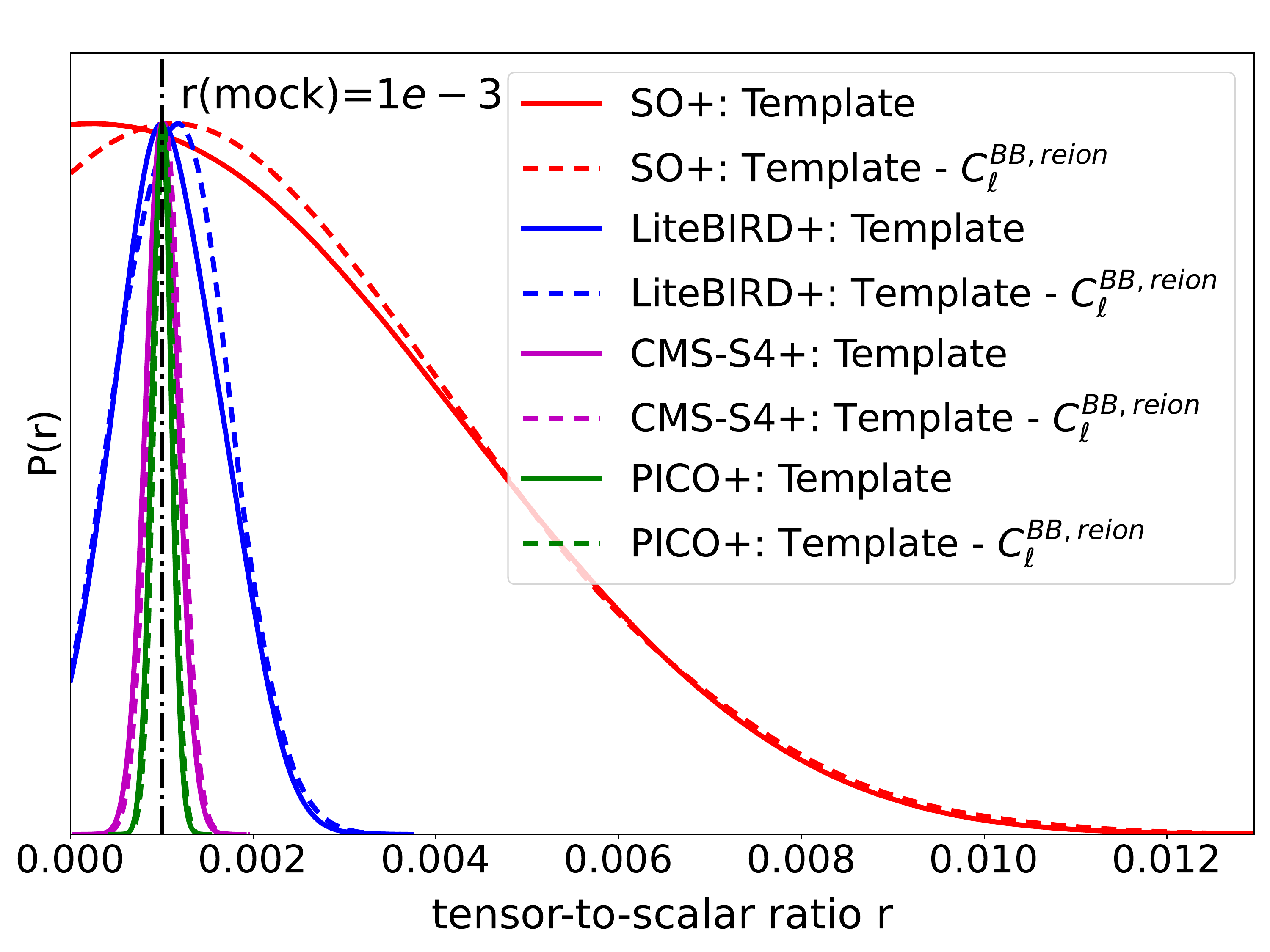}
    \caption{The marginalized posterior distribution for $r$ obtained from MCMC-based analysis for models Template (solid lines) and Template - $\CBB{\rm reion}$ (dashed lines) corresponding to different observatories have been presented.} 
    \label{fig:BBpostr}
\end{figure}

We infer the parameter $r$ for models of Template and Template $ - \CBB{\rm reion}$ for different cases. The inferred $r$ and bias $\Delta r / \sigma$ are shown in the Table~\ref{tab:r_estimates_nonoise}. Additionally, for reference, we have shown the posteriors of $r$ for input $r$ of $10^{-3}$ in Figure \ref{fig:BBpostr}. 

From Table~\ref{tab:r_estimates_nonoise}, the first obvious point to note is that the error bars on the measurement of $r$ also tighten with increased sensitivity and consequent improvement in the instrumental noise. We are more interested in the effect of reionization of inferred $r$, so to this end, we note that for the choice of model Template $- \CBB{\rm reion}$ the mean of the parameter $r$ is always overestimated when compared to the estimate of $r$ obtained for the model Template. This increase in the inferred mean value is intuitive as the model neglecting the patchy $B$-mode contribution has to compensate with a higher estimate of the parameter $r$. 

When comparing the response of the observatories to the two models Template and Template - $\CBB{\rm reion}$, the relevant quantity of interest is $\Delta r/\sigma$ a measure of the significance of the bias introduced as a result of incorrect modeling of $B$-mode power spectra. For a fiducial choice of $r=10^{-3}$, we find that LiteBIRD's inference on $r$ will suffer a bias of $\sim 0.03 \sigma$ but as the sensitivity of CMB experiments improves we may observe a bias of $\sim 0.2\sigma$ with space-based experiments like PICO. The bias is even more significant if the true value of $r$ is even lower at $5\times 10^{-4}$. In such a Universe,  even with ground-based experiments like CMB-S4 will always suffer an of $\sim 0.19\sigma$ and for PICO we will observe a bias of $\sim 0.23\sigma$. 

We present two more cases of bias estimation with CMB-S4 and PICO at higher delensing in the last two rows of Table \ref{tab:r_estimates_nonoise} at $95\%$ and $100\%$ for an input $r=5\times 10^{-4}$. While the $100\%$ delensing corresponds to a hypothetical case where we would have correctly reconstructed the lensing potential through the large-scale structure surveys, the $95\%$ delensing corresponds to the best delensing possible allowed by instrumental beam and sensitivity of PICO \citep{diego2020comparison}. As our ability to delens improves, our inference of $r$ becomes even more susceptible to confusion from patchy $B$-mode signal. With $95\%$ delensing the bias estimate for PICO is $\sim 0.33\sigma$ while for $100\%$ delensing\footnote{The case with $100\%$ delensing is a hypothetical scenario considered in the analysis to show the maximum impact on the $B$-mode signal.} it increases to $\sim 1.59\sigma$. For CMB-S4, $95\%$ and $100\%$ delensing translates to bias of $\sim 0.27 \sigma$ and $\sim 0.41\sigma$. We thus conclude that, for a given reionization history, the bias on $r$ will increase when the true value of $r$ is smaller and/or when the sensitivity of the instrument is better and/or when the delensing is more efficient.

The mock $B$-mode signal used in the above analysis is a prediction for the data to be observed in the future. However, in a realistic observing scenario, the data points will be affected by random errors arising from the telescope noise which the simulated mock does not capture. To incorporate these uncertainties in the likelihood, we add Gaussian random noise to the mock spectrum, the bias estimates corresponding to which have been presented in Appendix~\ref{app:grn_restimates}. We find that the bias estimate in such a case is consistent with the estimates presented above.

\begin{table}
    \centering
    \caption{Constraints on parameter $r$ presented as $\cc{{\bar{r}}^{\sigma_+}_{\sigma_-}}\times 10^3$ obtained from the MCMC analysis of models Template and Template - $\CBB{\rm reion}$ corresponding to different observatory cases have been presented. Here, $\bar{r}$ refers to the mean of the $r$ posterior while $\sigma_+$ and $\sigma_-$ refer to the $68\%$ limits of the posterior. The constraints for the two choices of the mock value of $r$ used in this analysis are $10^{-3}$ and $5\times 10^{-4}$ presented separately in this Table. $\Delta r/\sigma$ is a measure of the significance of the bias with respect to the error on the measurement and is presented in the fourth column.}
    \begin{tabular}{lccc}
    \hline
    Observatory case     & Model &  68$\%$ limits & $\Delta r/\sigma$\\
    \hline
    $10^3 \times r= 1$ &&&\\
    \multirow{2}{1em}{SO+}     & Template & $<3.77$ & \multirow{2}{2em}{$-$}\\[0.1cm]
    & Template - $\CBB{\rm reion}$ & $<3.84$ &\\[0.21cm]
    \multirow{2}{1em}{LiteBIRD+}     & Template & \pc{1.064}{0.497}{0.640} & \multirow{2}{2em}{0.027}\\[0.1cm]
    & Template - $\CBB{\rm reion}$ & \pc{1.079}{0.506}{0.633} &\\[0.21cm]
    \multirow{2}{1em}{CMBS4+}     & Template & \pc{0.999}{0.187}{0.188} & \multirow{2}{2em}{0.181}\\[0.1cm]
    & Template - $\CBB{\rm reion}$ & \pc{1.033}{0.188}{0.188} &\\[0.21cm]
    \multirow{2}{1em}{PICO+}     & Template & \pc{0.998}{0.106}{0.105} & \multirow{2}{2em}{0.208}\\[0.1cm]
    & Template - $\CBB{\rm reion}$ & \pc{1.020}{0.106}{0.106} &\\[0.21cm]
    \hline
    $10^3 \times r=0.5$&& &\\
    \multirow{2}{1em}{CMBS4+}     & Template & \pc{0.501}{0.184}{0.183} & \multirow{2}{2em}{0.185}\\[0.1cm]
    & Template - $\CBB{\rm reion}$ & \pc{0.535}{0.183}{0.183} &\\[0.21cm]
    \multirow{2}{1em}{PICO+}     & Template & \pc{0.499}{0.097}{0.097} & \multirow{2}{2em}{0.231}\\[0.1cm]
    & Template - $\CBB{\rm reion}$ & \pc{0.522}{0.097}{0.097} &\\[0.21cm]
    \hline
    $10^3 \times r=0.5$&Delensing at $95\%$ & &\\
    \multirow{2}{1em}{CMBS4+}     & Template & \pc{0.501}{0.125}{0.124} & \multirow{2}{2em}{0.272}\\[0.1cm]
    & Template - $\CBB{\rm reion}$ & \pc{0.535}{0.123}{0.123} &\\[0.21cm]
    \multirow{2}{1em}{PICO+}     & Template & \pc{0.500}{0.066}{0.065} & \multirow{2}{2em}{0.333}\\[0.1cm]
    & Template - $\CBB{\rm reion}$ & \pc{0.522}{0.065}{0.066} &\\[0.21cm]
    \hline
    $10^3 \times r=0.5$&Delensing at $100\%$ & &\\
    \multirow{2}{1em}{CMBS4+}     & Template & \pc{0.501}{0.092}{0.091} & \multirow{2}{2em}{0.413}\\[0.1cm]
    & Template - $\CBB{\rm reion}$ & \pc{0.539}{0.091}{0.091} &\\[0.21cm]
    \multirow{2}{1em}{PICO+}     & Template & \pc{0.500}{0.017}{0.017} & \multirow{2}{2em}{1.588}\\[0.1cm]
    & Template - $\CBB{\rm reion}$ & \pc{0.527}{0.017}{0.017} &\\[0.21cm]
    \hline
    \end{tabular}
    \label{tab:r_estimates_nonoise}
\end{table}

\subsection{Estimation of bias in $r$:  max-BB reionization model}
\label{subsec:maxBB}

The best-fit model of reionization, as inferred from the present CMB observations, may not necessarily correspond to the true model of reionization of the Universe. In order to study a case of extreme bias, we choose a model of reionization corresponding to the maximum allowed $B$-mode power spectra contribution allowed by 3$\sigma$ contours of the MCMC chains of the Planck+SPT case. This model was introduced in Section \ref{subsec:ksztauparam} and the corresponding reionization history can be found in Figure \ref{fig:ionizationfrac}. Corresponding to the model of reionization  $D^{BB,\mathrm{reion}}_{\ell=200}$ contribution is $18.41 \rm{n K}^2$ (for comparison, $D^{BB,\mathrm{reion}}_{\ell=200} = 7.03 \rm{n K}^2$ for the fiducial model). The bias estimate for the max-BB model of reionization for the case of CMBS4+ and PICO+ with an input $r=[1\times 10^{-3},5\times 10^{-4}]$ has been presented in Table \ref{tab:r_estimates3sigma_nonoise}. With an increased contribution from reionization, we find that for both choices of input $r$ we obtained a higher bias as expected. With $95\%$ delensing we begin to see the bias increase to $\gtrsim 0.5\sigma$ for both the choice of experiments. 

\begin{table}
    \centering
    \caption{Same as Table \ref{tab:r_estimates_nonoise} but with max-BB model of reionization in the mock $B$-mode power spectra.}
    \begin{tabular}{lccc}
    \hline
    Observatory case     & Model &  68$\%$ limits & $\Delta r/\sigma$\\
    \hline
     $10^3 \times r=1.0$&Delensing at $85\%$ &   &\\
    \multirow{2}{1em}{CMBS4+}     & Template & \pc{1.012}{0.186}{0.187} & \multirow{2}{2em}{0.316}\\[0.1cm]
    & Template - $\CBB{\rm reion}$ & \pc{1.071}{0.187}{0.187} &\\[0.21cm] 
    \multirow{2}{1em}{PICO+}     & Template & \pc{1.008}{0.106}{0.105} & \multirow{2}{2em}{0.368}\\[0.1cm]
    & Template - $\CBB{\rm reion}$ & \pc{1.047}{0.106}{0.106} &\\    
    \hline
     $10^3 \times r=1.0$&Delensing at $95\%$ &   &\\
    \multirow{2}{1em}{CMBS4+}     & Template & \pc{1.009}{0.132}{0.131} & \multirow{2}{2em}{0.537}\\[0.1cm]
    & Template - $\CBB{\rm reion}$ & \pc{1.080}{0.132}{0.132} &\\[0.21cm] 
    \multirow{2}{1em}{PICO+}     & Template & \pc{1.010}{0.072}{0.071} & \multirow{2}{2em}{0.639}\\[0.1cm]
    & Template - $\CBB{\rm reion}$ & \pc{1.056}{0.072}{0.072} &\\    
    \hline
     $10^3 \times r=0.5$&Delensing at $85\%$ &   &\\
    \multirow{2}{1em}{CMBS4+}     & Template & \pc{0.518}{0.184}{0.185} & \multirow{2}{2em}{0.368}\\[0.1cm]
    & Template - $\CBB{\rm reion}$ & \pc{0.586}{0.185}{0.185} &\\[0.21cm] 
    \multirow{2}{1em}{PICO+}     & Template & \pc{0.505}{0.097}{0.097} & \multirow{2}{2em}{0.433}\\[0.1cm]
    & Template - $\CBB{\rm reion}$ & \pc{0.547}{0.098}{0.097} &\\    
    \hline
     $10^3 \times r=0.5$&Delensing at $95\%$ &   &\\
     
    \multirow{2}{1em}{CMBS4+}     & Template & \pc{0.514}{0.126}{0.125} & \multirow{2}{2em}{0.555}\\[0.1cm]
    & Template - $\CBB{\rm reion}$ & \pc{0.584}{0.126}{0.126} &\\[0.21cm]       
    \multirow{2}{1em}{PICO+}     & Template & \pc{0.508}{0.066}{0.065} & \multirow{2}{2em}{0.727}\\[0.1cm]
    & Template - $\CBB{\rm reion}$ & \pc{0.556}{0.066}{0.066} &\\    
    \hline    
    \end{tabular}
    \label{tab:r_estimates3sigma_nonoise}
\end{table}

\subsection{Caution for $5\sigma$ measurement of tensor-to-scalar ratio by upcoming CMB missions}\label{sec:discussion}

The stage-4 CMB experiments are targeting a $5\sigma$ measurement of the tensor to scalar ratio $r$. A huge effort is underway to observe the pristine primordial $B$-mode power spectra by mitigating the instrument noise, galactic foregrounds, and extra-galactic foregrounds. One of the key aspects which makes it possible to distinguish between CMB and foregrounds is their distinguishable frequency spectrum between a few tens of GHz to nearly THz frequency range. However, the effect of patchy reionization which is discussed in this paper leads to the same kind of spectrum as CMB, hence cannot be distinguished from the actual CMB signal using the frequency spectrum. This makes the extragalactic contamination due to patchy reionization particularly critical. As the model of reionization is not well known and large fluctuations in the electron density during reionization can cause large fluctuations in $B$-mode polarization, we need to make sure that the inferred value of tensor to scalar ratio is due to primordial gravitational waves and not due to patchy reionization. This is significant in particular as CMB measurements are the only observational probes to measure primordial gravitational waves over these frequencies.

In this work, we comprehensively studied the nature of this contamination on tensor to scalar ratio for different scenarios of reionization and formulated a framework that can self-consistently mitigate the contamination from patchy reionization by combining multiple CMB probes such as E-mode polarization, kSZ, and $B$-mode polarization. The joint study enables to construct of a \textit{data-driven} model of reionization and can make it feasible to obtain an upper bound (or a measurement) of patchiness in electron density during the epoch of reionization and hence can also provide an upper bound (or measurement) of the $B$-mode signal arising from this effect.

Using a self-consistent analysis framework proposed in this work, it is quite evident that with planned sensitivities of $B$-mode observations by CMB-S4 and PICO we will observe a bias of $>0.15\sigma$ in the inference in $r$. For the allowed scenarios of reionization that are consistent with the kSZ temperature fluctuation detected by SPT \citep{Reichardt2020},  neglecting the reionization contribution can lead to a $\sim 18\%$ bias with respect to the standard deviation from CMB-S4 when $r = 10^{-3}$. We can generalize this inference that a $5\sigma$ detection with CMB-S4 would rather be a $\sim 4.82\sigma$ detection when the patchy $B$-mode is neglected. This scenario worsens if the true $r$ were smaller in such a case the contribution of reionization will become even more dominant. 

Following the same argument, a probable $5\sigma$ detection with a more sensitive probe like PICO would rather be a $\sim 4.8 \sigma$ detection. Additionally, for completeness, we chose a model of reionization allowed by current CMB measurements with maximum possible patchy $B$-mode signal amplitude. In such a case we find that for a Universe with $r=5\times 10^{-4}$, we will achieve a maximum bias of $\sim 0.56\sigma$ with CMB-S4 and $\sim 0.73\sigma$ with PICO further affecting our chance of claiming the detection of the actual value. In an extrapolation to our results, we additionally would want to claim that when and if more sensitive experiments come up and if true $r$ than $5\times 10^{-4}$ were even lower, dealing with patchy reionization will be challenging and to improve our understanding of patchy $B$-modes any further than this analysis would require redshift based information of galaxy evolution information which the line-of-sight integrated CMB observables lack. We also learned that $B$-mode observations even with the choice of the correct model may not help with exploring reionization any more than we will learn with the optical depth $\tau$ and kSZ measurement due to the monotonic shape of the power spectrum of $C_\ell^{BB}$ with the angular multipoles $\ell$ for scales larger than about a degree and degeneracy with other contamination such as lensing.

While, on one side of the coin we have been discussing how the bias may impact future observations of $r$, on the flip side, a bias of such high significance also points to the opportunity of making an independent detection of patchy $B$-mode signal from reionization by the upcoming CMB telescopes using the correlation between different angular multipoles $\ell$. This detection will be complementary to other upcoming probes of reionization, e.g., the 21~cm emission \citep{mellema2013reionization,mellema2015hi,choudhury2016modelling}.  We would like to explore such a possibility with the upcoming $B$-mode by observing experiments of CMB-S4 and PICO in our future work using simulations.

In our study, we assume that the galactic foreground components will be effectively addressed through significant efforts in the future. In this regard, with the current understanding of the foreground model, both parametric and blind component separation techniques have been carried out for the upcoming CMB experiments (SO, \cite{Ade_2019}; LiteBIRD, \cite{allys2022probing}; CMB-S4, \cite{abazajian2022cmb}; PICO, \cite{aurlien2022foreground}). It was found that while SO, LiteBIRD, and CMB-S4 would incur a foreground residual bias of $\lesssim \sigma(r)$, $\sigma(r)$ being the pessimistic detection sensitivity of the experiment, PICO would make an unbiased estimate of $r$. Depending on the choice of the sky model and component separation technique, the patchy reionization bias may appear to be sub-dominant when compared to foreground residual bias. But with techniques like multi-tracer de-lensing \citep{namikawa2022simons} and wideband  galactic foreground studies \citep{basu2019cmb}, it is expected that foreground bias would further reduce in the future.  So, as we move towards increasingly sensitive experiments, an improved sky model will demand accountability of the patchy $B$-mode contribution from reionization in the total $B$-mode spectra for an unbiased detection of $r$.

Before concluding this section, let us put our results in the context of other results in this area.  As far as reionization modelling is concerned, the large-scale $B$-mode spectra can be computed using a relatively simplistic model, e.g., the spherical bubble-based prescription of patchy reionization  \citep{baumann2003small,mortonson2007maximum,Dvorkin:2008tf,roy2018observing} and also the more rigorous numerical simulation of patchy reionization as employed in this work and \citet{mukherjee2019patchy,10.1093/mnrasl/slaa185,roy2021revised}. The simplistic spherical bubble-based prescriptions become inaccurate when individual ionized bubbles begin to overlap, thus, our estimates of $B$-mode spectra are more robust. Apart from \cite{10.1093/mnrasl/slaa185} and this work, we are not aware of any study that discusses constraints on $D^{BB,\text{reion}}_{\ell}$ at large scales using a physical model of reionization based on recent measurements of reionization observables. \cite{roy2021revised} predicted $D^{BB,\text{reion}}_{\ell=100}$ at $4 ~ \mathrm{nK}^2$ using a physical model of reionization assuming that  haloes with $M_{\text{min}}>10^9 M_\odot$ contribute to the ionization process, however, a one-to-one comparison between their results and our results is difficult due to their inherent assumption that ionizing photons from atomic cooled haloes do not contribute to reionization.

We also want to emphasize that this is the first study to capture the patchy reionization bias introduced in $r$ within a self-consistent Bayesian framework through joint estimations of different reionization observables. Previous studies \citep{mukherjee2019patchy,roy2021revised} attempted to estimate the contribution from patchy reionization based on data of Thomson scattering optical depth $\tau$ or Ly$\alpha$ data and estimated bias on $r$ based on Fisher framework. \cite{mukherjee2019patchy},  using \texttt{SCRIPT}, generated ionization maps for models of reionization consistent with Planck constraints on $\tau$ \citep{PlanckCollaboration2018}, observed that if contamination from patchy reionization is neglected in the analysis of $B$-mode polarization data, a maximum bias of about $30\%$ in the value of $r = 10^{-3}$ would be obtained. Similar conclusions were  obtained in the study by \cite{roy2021revised} where they used radiative transfer simulations calibrated to Ly$\alpha$ data to model their reionization history. Under the Bayesian framework, the forecasts for patchy reionization bias, we obtain, are consistent with the prediction of previous studies.

In a comparatively earlier work, \cite{baumann2003small}  used the halo approach \citep{komatsu2002sunyaev} to model the electron density distribution in the reionization era and calibrated their model using COBE observations at large scales. Assuming a delensing at $90\%$ they had commented that, the patchy $B$-mode from reionization would constitute a background for models of inflation with energy scale around $E_{\text{infl}}\lesssim 10^{15}$~GeV. An $r=5\times 10^{-4}$ corresponds to an energy scale of $\sim 5 \times 10^{15}$~GeV and through our study, we find fractional biases are indeed present when estimating $r$. Our forecasts portray a reionization bias which is in the ballpark with the predictions in \cite{baumann2003small}.

\section{Conclusion}\label{sec:conclude}

Sensitive observations of the $B$-mode polarization signal of the CMB in the future would enable the first detection of tensor-to-scalar ratio parameter $r$. This would help us to reject a large class of inflationary models, enabling us to gain insights into the mechanisms that laid the seeds of structure formation.
Unbiased detection of $r$ is hence of critical importance. In this effort, we attempted to estimate the bias that might be introduced in the inference of $r$ when and if the contribution from patchy $B$-modes is neglected. We propose a framework that can mitigate the contamination from patchy reionization on tensor to scalar ratio by estimating an upper bound (or a measurement) on it by using a \textit{data-driven} model of reionization constructed using different observables to reionization such as optical depth and kSZ signal.

If the patchy reionization effect is not included, then for the allowed range of reionization models considered in this analysis,  we find that with ground-based observatories like CMB-S4, the inference on $r$ would be biased at $>0.17\sigma$. This bias becomes more significant when PICO starts $B$-mode observations. We showed a case of extreme bias of $\sim 0.73\sigma$ which observations with PICO might experience with aggressive delensing of $95\%$. Further, we showed how this bias may impact the claim of the $5\sigma$ detection of $r$ by the Stage-4 CMB experiments if reionization is significantly inhomogeneous. While exploring the bias in $r$ we additionally found that even with the correct choice of model of $B$-mode the constraints on reionization parameters were governed by the sensitivity of $\tau$ and kSZ measurement. However, one important caveat that needs to be kept in mind is that the expected bias which is estimated depends on the models of reionization and the current constraints from Planck \citep{PlanckCollaboration2018} and SPT \citep{Reichardt2020}. In reality, reionization can be far more complicated and the impact on the $B$-mode can be significant. As a result, the proposed technique needs to be implemented to make a robust interpretation of an observed $B$-mode polarization signal.

The imprints of reionization can also lead to additional correlations between different angular multipoles for temperature, and polarization. This can potentially further improve our understanding during the epoch of reionization. In the future, we would like to explore the possibility of measuring such signatures beyond the power spectrum. Also, in addition, 21-cm measurements will open a complementary probe to understand the epoch of reionization, which can help in further understanding the history of reionization and the morphology of the inhomogeneities in electron density during reionization. This will further improve in limiting any contamination from patchy reionization on $B$-mode polarization and can measure primordial gravitational wave signal immune from at least one extra-galactic foreground contamination having the same frequency spectrum as CMB.

\section*{Acknowledgements}
DJ and TRC acknowledge support of the Department of Atomic Energy, Government of India, under project no. 12-R\&D-TFR-5.02-0700. The work of SM is a part of the $\langle \texttt{data|theory}\rangle$ \texttt{Universe-Lab} which is supported by the TIFR and the Department of Atomic Energy, Government of India. 

\section*{Data Availability}

A basic version of the semi-numerical code \texttt{SCRIPT} for generating the ionization maps used in the paper is publicly available at \url{https://bitbucket.org/rctirthankar/script}. Any other data related to the paper will be shared on reasonable request to the corresponding author (DJ).



\bibliographystyle{mnras}
\bibliography{main} 




\appendix

\section{Missing Power in momentum field for finite sized simulation boxes}\label{app:mispower}

\begin{figure}
	\includegraphics[width=\columnwidth]{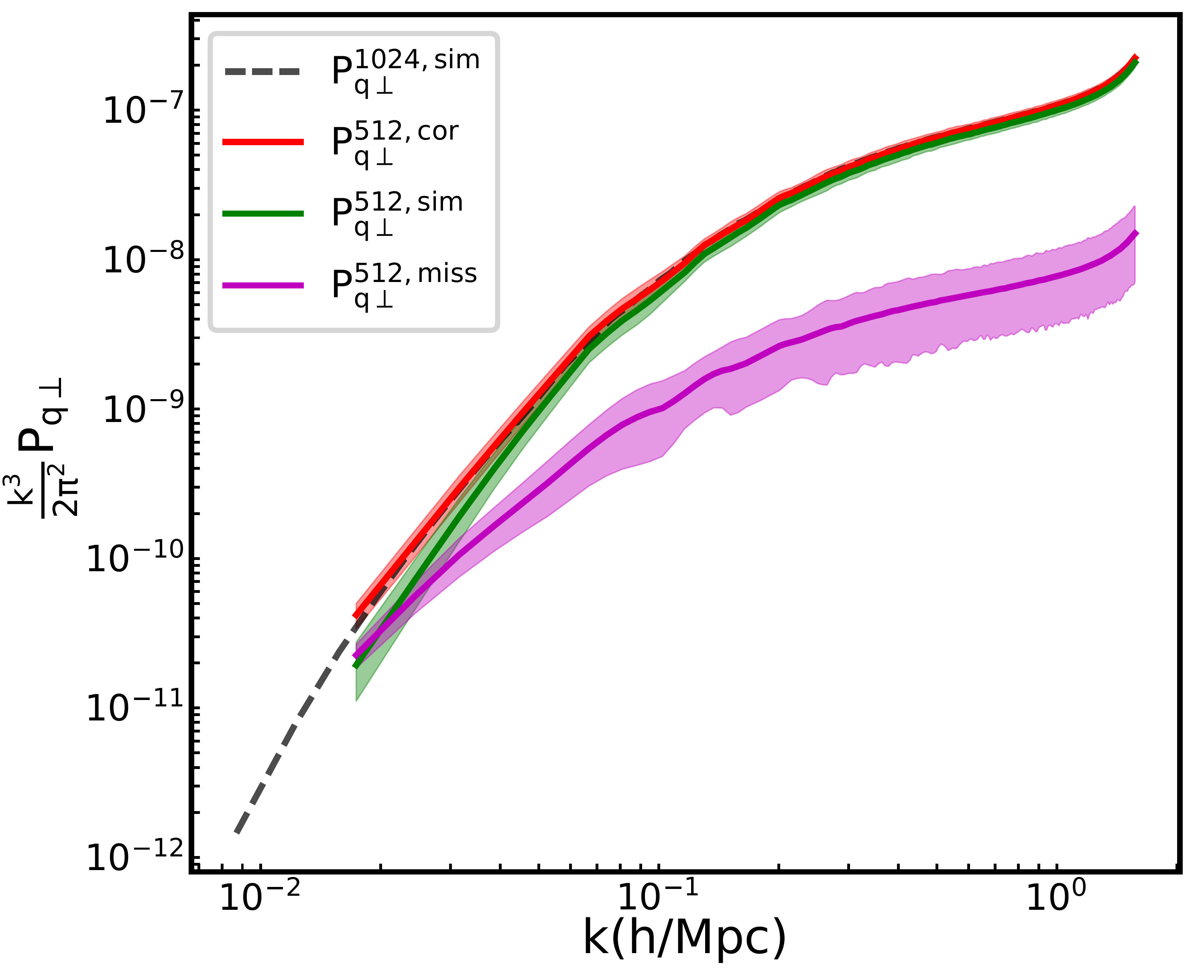}
    \caption{Power in transverse component of ionized momentum field for a box of $512 ~ h^{-1}$ Mpc for cases considering (in red) and neglecting (in green) the contribution of missing power in ionized momentum field arising corresponding to wave mode range of $2\pi /1024 h^{-1}~\text{Mpc}\leq k \leq 2\pi/512h^{-1}~\text{Mpc}$ is shown here. Here, $P^{512, \mathrm{sim}}_{q_\perp}$ and $P^{1024, \mathrm{sim}}_{q_\perp}$ refers to the power in transverse component of momentum field evaluated from the simulation box of length of 512 and 1024 $h^{-1}$ Mpc while $P^{512, \mathrm{miss}}_{q_\perp}$ refers to the missing power evaluated through Equation \eqref{eq:P_qmisfin}. The corrected power is then evaluated as $P^{512, \mathrm{cor}}_{q_\perp}=P^{512, \mathrm{sim}}_{q_\perp}+P^{512, \mathrm{miss}}_{q_\perp}$. In all curves, the shaded region corresponds to the standard deviation in power after it has been averaged over several realizations.} 
    \label{fig:corr512}
\end{figure}

\begin{figure}
	\includegraphics[width=\columnwidth]{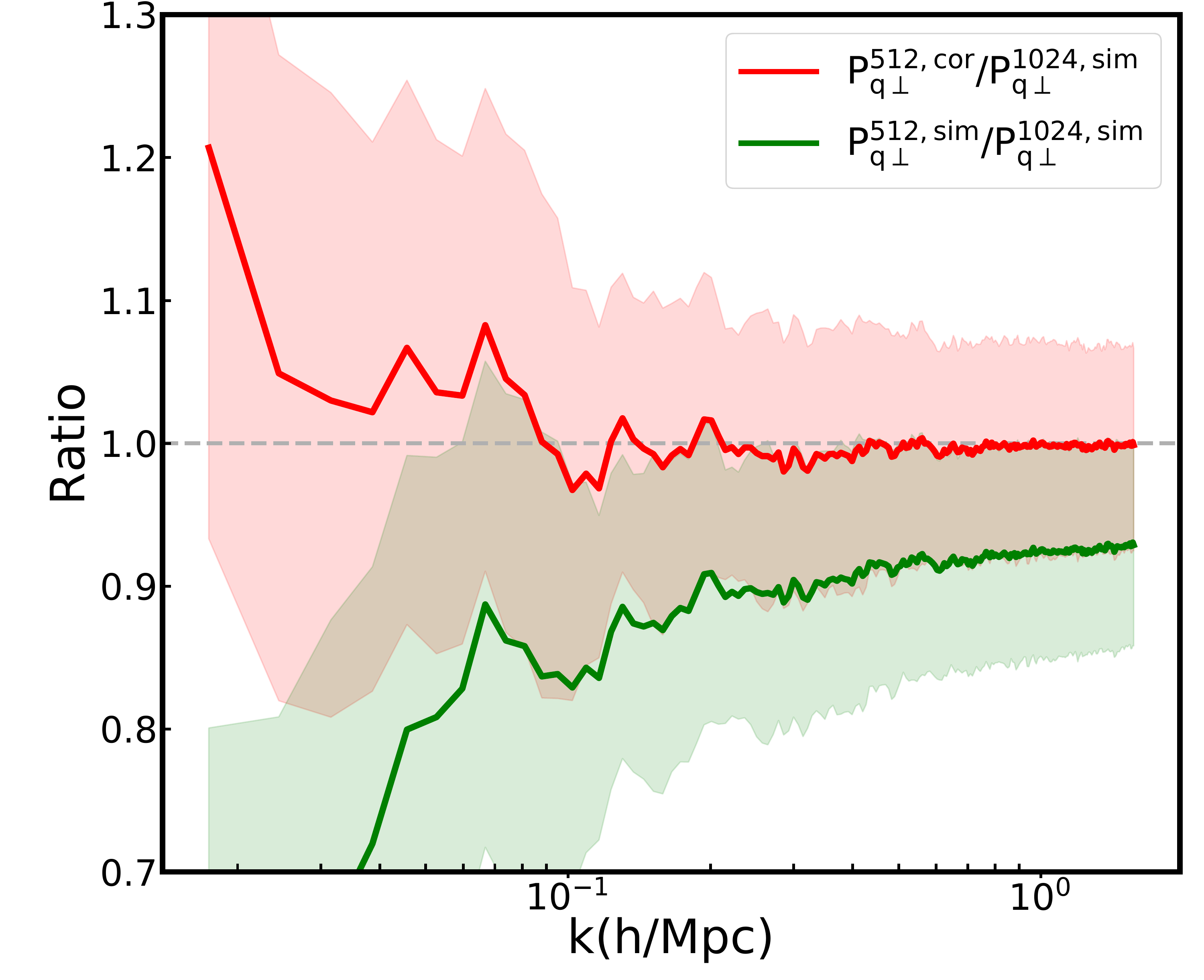}
    \caption{Ratio of $P^{512, \mathrm{cor}}_{q_\perp}$ and $P^{512, \mathrm{sim}}_{q_\perp}$ with respect to $P^{1024, \mathrm{sim}}_{q_\perp}$ is shown here in order to infer convergence of $P_{q_{\perp}}$.} 
    \label{fig:corrcon512}
\end{figure}

\begin{figure}
	\includegraphics[width=\columnwidth]{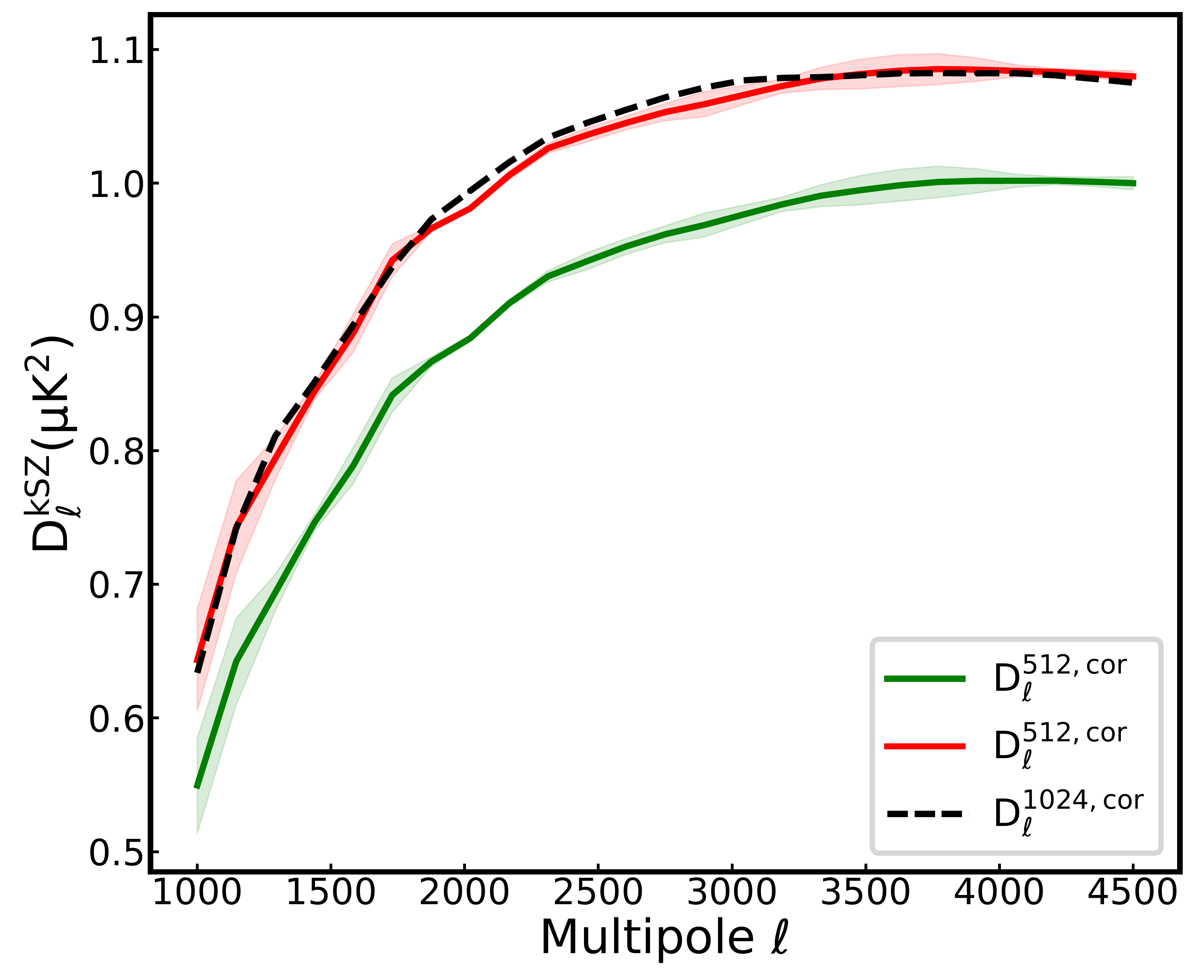}
    \caption{Convergence of angular power spectra of kSZ evaluated for a simulated and corrected box of length $512 ~ h^{-1}$ Mpc with respect to angular power spectra of a simulated box of length $1024 ~ h^{-1}$ Mpc is shown.} 
    \label{fig:ksz512}
\end{figure}

\begin{figure}
	\includegraphics[width=\columnwidth]{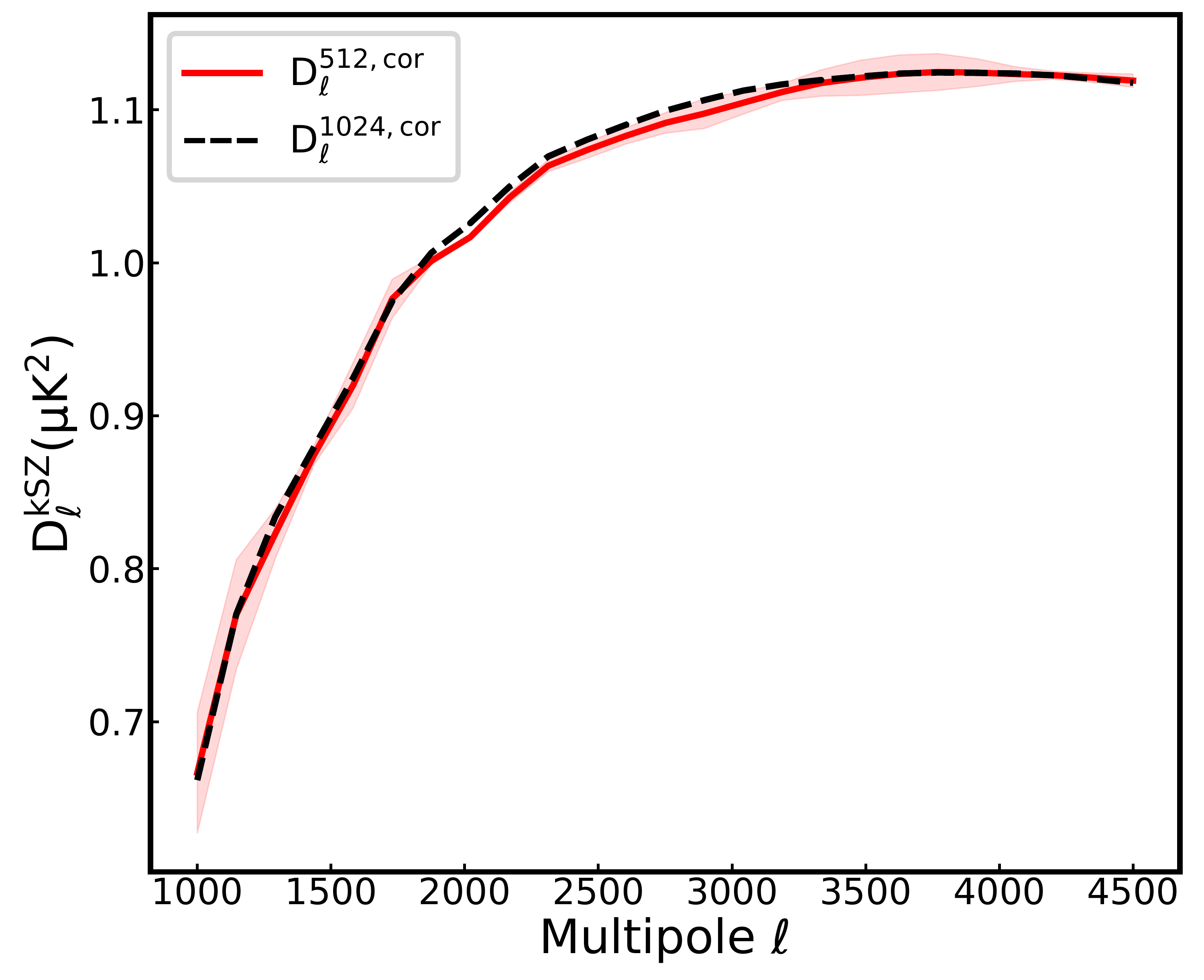}
    \caption{Convergence of corrected angular power spectra of kSZ evaluated for a box of length $512 ~ h^{-1}$ Mpc (in red) with respect to corrected angular power spectra for a box of length $1024 ~ h^{-1}$ Mpc (in dashed) is shown. Here, power in $P_{q_\perp}$ has been corrected by considering velocity modes ranging from $2\pi/8192 h^{-1}~\text{Mpc}  \leq k \leq 2\pi/L_\mathrm{box} h^{-1}~\text{Mpc} $. } 
    \label{fig:allcorr512}
\end{figure}

The transverse component of the momentum field power spectrum is given as \citep{Park_2013}
\bear
P_{q_\perp}(k,z)&=\int \frac{d^3k^\prime}{(2\pi)^3}(1-{\mu^\prime}^2 )\Big[ P_{ee}(|\mbf{k}-\mbf{k^\prime}|)P_{vv}(k^\prime)\\
&-\frac{k^\prime}{|\mbf{k}-\mbf{k^\prime}|}P_{ev}(|\mbf{k}-\mbf{k^\prime}|)P_{ev}(k^\prime)\Big],
\ear
where $P_{vv}(k)$ is the velocity power spectrum and $P_{ev}$ is the cross power spectrum of the ionization and velocity fields. The missing power arises as we do not have information about the power spectra $P_{ee}(k)$, $P_{vv}(k)$, $P_{ev}(k)$ for $k<k_{\mathrm{box}}\equiv 2\pi/L_{\mathrm{box}}$ where $L_{\mathrm{box}}$ is the length of the simulation box. As argued by \cite{Park_2013}, for $k^\prime < k_{\mathrm{box}}$, most of the missing power is contributed by $P_{ee}(|\mbf{k}-\mbf{k^\prime}|)P_{vv}(k^\prime)$ because of the way the velocity field scales at large scales. The missing power is thus given by
\begin{equation}
    \begin{aligned}
    &P^{missing}_{q_\perp}(k,z)\\
    & = \int_{k^\prime<k_{box}} \frac{d^3k^\prime}{(2\pi)^3}(1-{\mu^\prime}^2)P_{ee}(|\mbf{k}-\mbf{k}^\prime|,z)P_{vv}(k^\prime,z),
    \end{aligned}
\end{equation}
where $\mu^\prime=\hat{\mbf{k}}\cdot\hat{\mbf{k}}^\prime$.

It is possible to calculate the expected missing power analytically at very large scales. At scales larger than the bubble sizes, we know that the electron density power spectrum is a scaled version of dark matter power spectra given as $P_{ee}(k,z)=\mathcal{R}^2(k,z)D^2(z)P_{\delta\delta}(k)$ \citep{mukherjee2019patchy} where $\mathcal{R}(k,z)\equiv \chi_{He}Q_{HII}(z)b_h(k,M_{min},z)$ and $D(z)$ is the linear growth factor. At very large scales $\lim_{k \to 0} b_h(k,z)=b(z)$ and hence we denote $\mathcal{R}(z)=b(z)\chi_{He}Q_{HII}(z)$.

Further, in the linear regime, we can relate the velocity power spectrum to the density power spectrum as
\begin{equation}
    P_{vv}(k)=\cc{\frac{f\Dot{a}}{k}}^2P_{\delta\delta}(k)
    \label{eq:Pvv}
\end{equation}
where, $f = d \log D / d \log a$. The missing power is then
\begin{equation}
    \begin{aligned}
    P^{missing}_{q_\perp}(k,z)&={\mathcal{R}}^2(z)D^2(z)\cc{\frac{f\Dot{a}}{(2\pi)^{3/2}}}^2\\&\int_{k^\prime<k_{box}} \frac{d^3k^\prime}{{k^\prime}^2}(1-{\mu^\prime}^2)P_{\delta\delta}(|\mbf{k}-\mbf{k}^\prime|)P_{\delta\delta}(k^\prime),\\
    \end{aligned}
    \label{eq:P_qmisfin}
\end{equation}
which can be computed analytically.

As an illustration we consider the momentum field power spectra for two boxes, one with length $L_{\mathrm{box}} = 512 h^{-1} \text{Mpc}$ and another with  $L_{\mathrm{box}} = 1024 h^{-1} \text{Mpc}$. We plot the corresponding power spectra $P^{512, \mathrm{sim}}_{q_\perp}(k)$ (green) and $P^{1024, \mathrm{sim}}_{q_\perp}(k)$ (black), uncorrected for the missing power, in Figure \ref{fig:corr512}. It is clear that at large scales $k \lesssim 0.03 h \text{Mpc}^{-1}$, the power spectra of the smaller box are lower than that of the larger box which is a clear indication of the missing modes. We then add the power contributed by missing velocity modes in the range $2\pi / 1024 h^{-1}~\text{Mpc} \leq k \leq 2\pi / 512 h^{-1}~\text{Mpc}$ using the Equation \eqref{eq:P_qmisfin} to $P^{512, \mathrm{sim}}_{q_\perp}(k)$, let us call the corrected power spectrum as $P^{512, \mathrm{cor}}_{q_\perp}(k)$. It is evident that the corrected power spectrum (red) agrees with the larger box quite well. The amplitude of the missing power is shown by the magenta curve and denoted here as $P^{512, \mathrm{miss}}_{q_\perp}(k)$. 

The consequential convergence of $P_{q_\perp}$ for box of length $512 h^{-1}$ Mpc with that for a box of length $1024 h^{-1}$ Mpc upon adding the missing power is prominent in Figure \ref{fig:corrcon512}. Here, we plot the ratio of  $P^{512, \mathrm{cor}}_{q_\perp}$ to  $P^{1024, \mathrm{sim}}_{q_\perp}$ (in red) and the ratio of  $P^{512, \mathrm{sim}}_{q_\perp}$ to  $P^{1024, \mathrm{sim}}_{q_\perp}$ (in green). We find that at $k \gtrsim 0.1$ the ratio corresponding to $P^{512,\mathrm{cor}}_{q_\perp}$ is close to 1 while that corresponding to $P^{512,\mathrm{sim}}_{q_\perp}$ is $\sim 0.9$. 

Finally, after correcting for missing power at each redshift corresponding to our model of reionization, we plot the kSZ angular power spectra corresponding to both the boxes as shown in Figure \ref{fig:ksz512}. We denote kSZ angular power spectra for $512h^{-1}$ Mpc box with corrected power as  $D^{512,\mathrm{cor}}_\ell$ (in red) while the other one as $D^{512,\mathrm{sim}}_\ell$ (in green). In the dashed line we denote the kSZ power corresponding to the simulation box of length $1024 h^{-1}$ Mpc. We find that post correcting for missing power the kSZ angular power for both the boxes is fairly convergent.

In this spirit, for our analysis throughout the paper we consider to correct for missing velocity modes ranging from $2\pi/8192 h^{-1}~\text{Mpc} \leq k \leq 2\pi/512 h^{-1}~\text{Mpc} $ to account for missing power in $P_{q_\perp}$ at even larger scales. In Figure \ref{fig:allcorr512} we show the convergence of kSZ angular power spectra evaluated for boxes with length $512$ and $1024 h^{-1}$ Mpc when missing power is corrected for wavemode range of $2\pi/8192 h^{-1}~\text{Mpc}  \leq k \leq 2\pi/L_\mathrm{box} h^{-1}~\text{Mpc} $.

\section{Applicability of the Limber approximation}\label{app:limber}

\begin{figure}
	\includegraphics[width=\columnwidth]{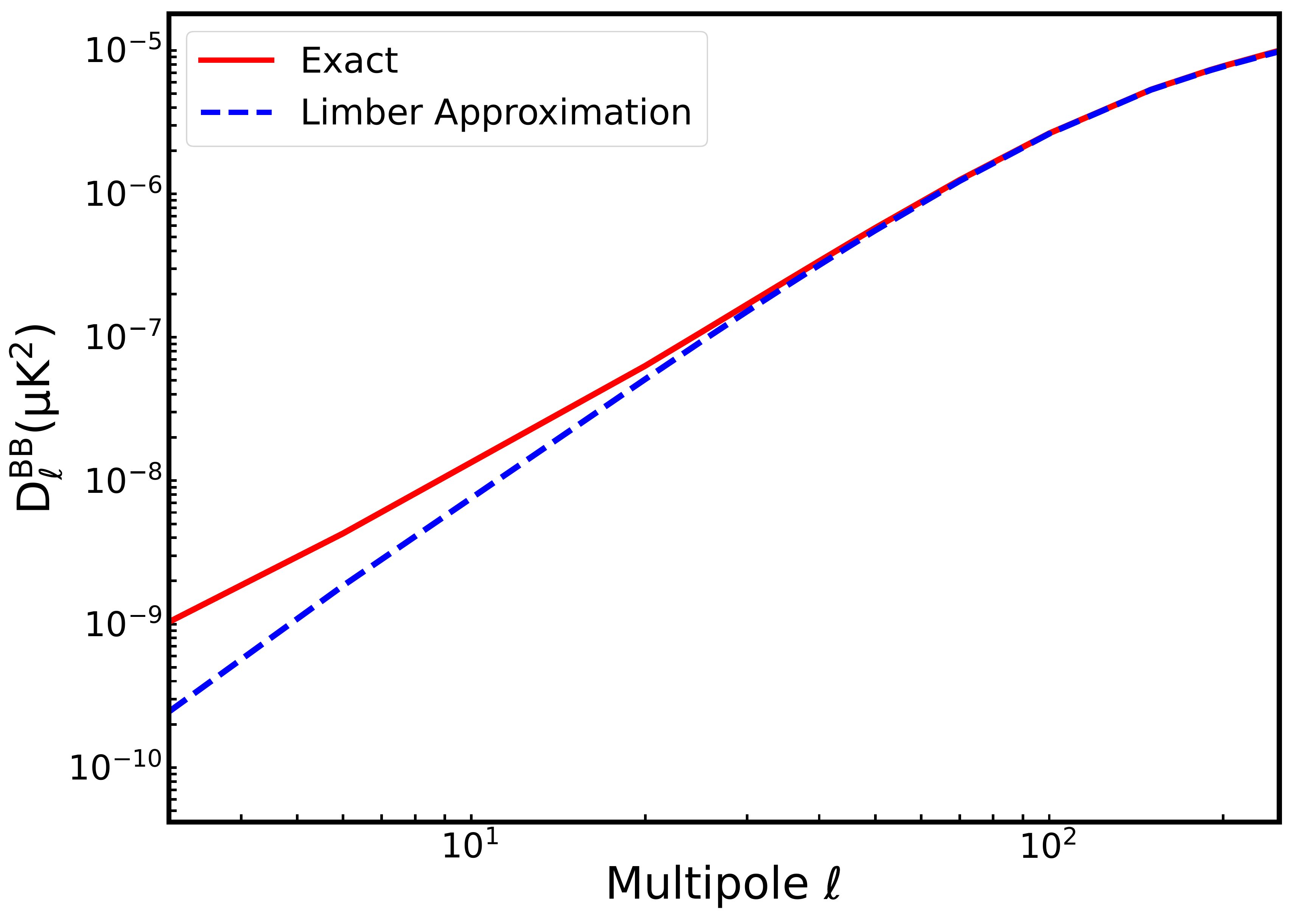}
    \caption{The $B$-mode angular power spectrum arising from patchy reionization has been shown for the exact method of evaluation using Equation \eqref{eq:BBexact} and the Limbers method of evaluation using Equation \eqref{eq:BB_lim}} 
    \label{fig:limber512}
\end{figure}

\begin{figure}
	\includegraphics[width=\columnwidth]{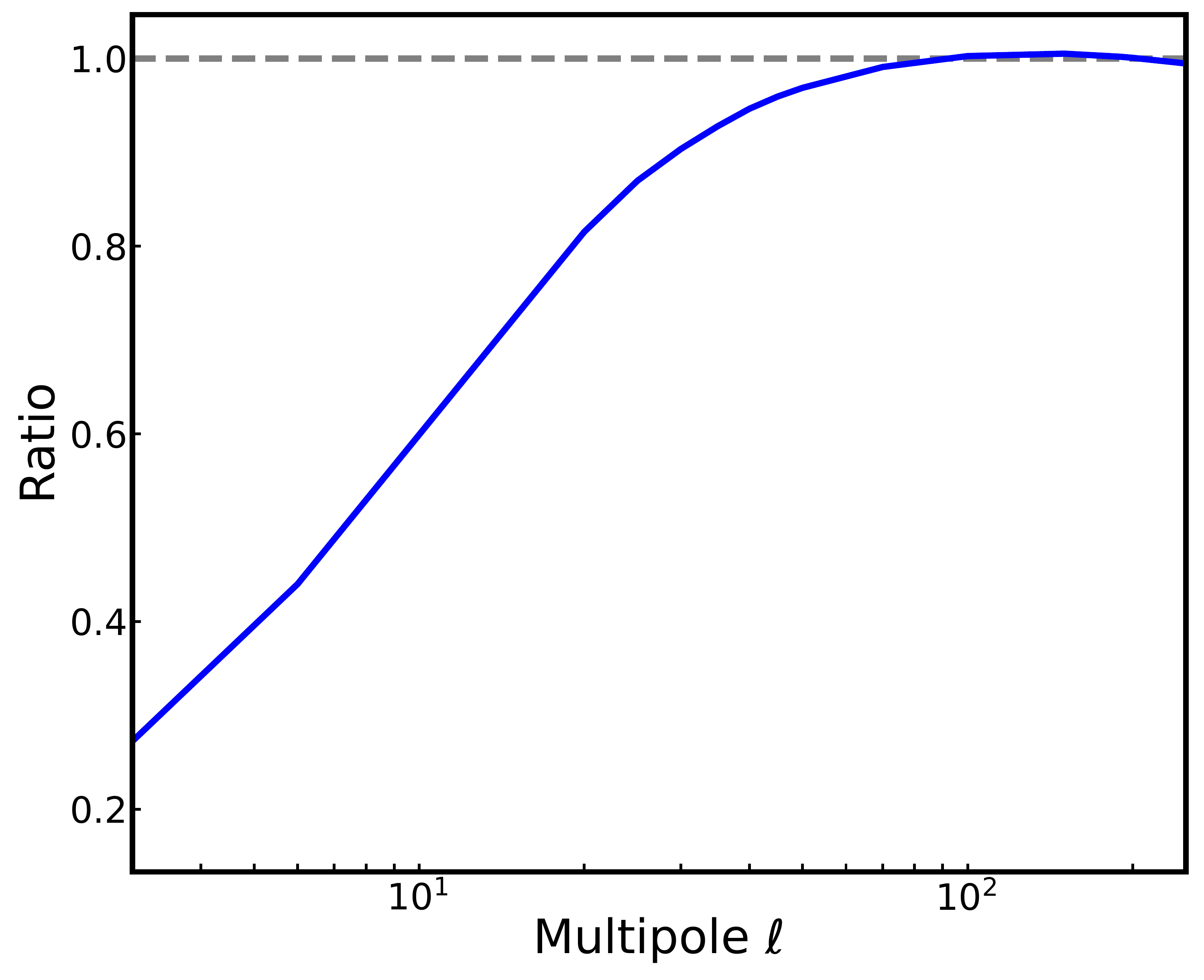}
    \caption{Ratio of $B$-mode power spectrum evaluated from Limber approximation to $B$-mode power evaluated from exact method has been shown.} 
    \label{fig:limber512ratio}
\end{figure}

The patchy $B$-mode angular power spectra in this study are evaluated under the assumption of Limber approximation with an intent to relax the complexity of the code during MCMC sampling. Let us discuss the validity of the approximation in more detail. In general, the $B$-mode arising from Thomson scattering of CMB photons with the patchy spatial distribution of free-electrons in the reionization era can be evaluated as
\begin{equation}
\begin{aligned}
C_\ell^{BB,\mathrm{  reion}}&=\frac{24\pi \bar{n}^2_{H}\sigma^2_T}{100} \int d\chi (1+z)^2\int d\chi' (1+z')^2e^{-\tau(\chi)- \tau(\chi')} \\
&\times \int dk \frac{k^2}{2\pi^2} P_{ee} (k, \chi, \chi') j_\ell(k\chi)j_\ell(k\chi') \frac{Q_{\rm RMS}^2}{ {2}}\\
\end{aligned}
\label{eq:BBexact}
\end{equation}
where the power spectrum of the electron fraction $x_e$ is written as $\langle x_e ({\mathbf{k}},\chi')  x_e^* ({\mathbf{k'}},\chi')\rangle  \equiv P_{ee} ({\mathbf{k}}, \chi, \chi') \delta_D({\mathbf{k}} - {\mathbf{k}}')$. Here, $j_\ell(k\chi)$ are the spherical Bessel functions. Therefore to evaluate even a single multipole $\ell$ of $C_\ell^{BB,\mathrm{reion}}$ corresponding to a model of reionization we ought to evaluate three integrals (i.e. over $d\chi, ~ d\chi^\prime, ~ dk$) over the redshift range of $5 \leq z \leq 20$. This is a computationally expensive step when considering the objective of Bayesian inference by sampling the free parameter space of reionization. In such a case, one incorporates Limber approximation and by using the single integral in Equation \eqref{eq:BB_lim} computes $C_\ell^{BB,\mathrm{  reion}}$ at a relatively cheaper computational load. In Figure \ref{fig:limber512} we show the $B$-mode angular power evaluated from the Limber approximation (shown in a blue curve) as well as the power evaluated from the exact method (shown in a solid red curve). We find that at multipole ($\ell \lesssim 30$) the angular power is under-estimated when the calculation is made under the Limber approximation. We ascertain this finding by plotting the ratio of the power evaluated via the two methods in Figure \ref{fig:limber512ratio}.

With regards to our bias estimate presented in Sections \ref{subsec:bmoderesults} and \ref{subsec:maxBB}, as we employ the Limber approximation to evaluate the $B$-mode angular power spectrum throughout our study, from the above discussion, we can infer that any bias estimate $\Delta r$ from our study should always be treated as a lower limit in principle.

\begin{figure}
	\includegraphics[width=\columnwidth]{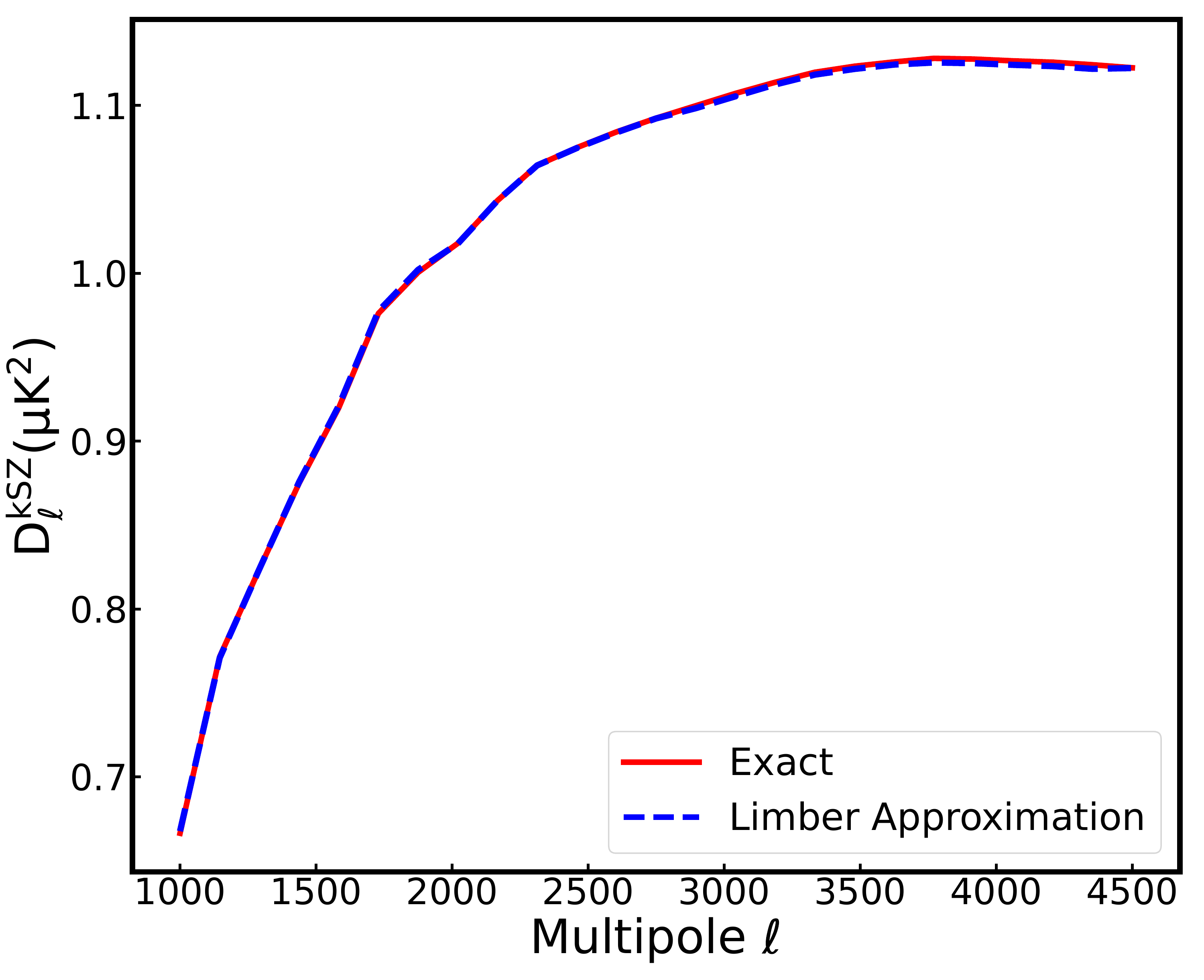}
    \caption{The kSZ angular power spectrum arising from patchy reionization has been shown for the exact method of evaluation using Equation \eqref{eq:fullkSZ} and the Limbers method of evaluation using Equation \eqref{eq:ksz_lim}.} 
    \label{fig:limberksz512}
\end{figure}

Let us now discuss the validity of the Limber approximation for the kSZ signal. The true kSZ angular power spectrum is given by \citep{Park_2013}
\begin{equation}
\begin{aligned}
    C&^{\mathrm{kSZ,reion}}_\ell=\frac{\ell(\ell+1)}{\pi}\cc{\sigma_T \bar{n}_{H}T_0}^2\int d\chi (1+z)^2e^{-\tau(\chi)}\\
    &\times \int d\chi^\prime(1+z^\prime)^2 e^{-\tau(\chi^\prime)}
    \int k^2 dk \frac{j_\ell(k\chi)}{k\chi} \frac{j_\ell(k\chi^\prime)}{k\chi^\prime}P_{q_\perp}(k,\chi)
    \label{eq:fullkSZ}
\end{aligned}
\end{equation}
where the power spectrum of  transverse component of the Fourier transform of momentum field $q_\perp(\mbf{k},z)$ is written as $\langle q_\perp ({\mathbf{k}},z)  q_\perp^* ({\mathbf{k'}},z)\rangle  \equiv P_{q_\perp} ({\mathbf{k}},z) \delta_D({\mathbf{k}} - {\mathbf{k}}')$ and $j_\ell(k\chi)$ are the spherical Bessel function.
Just like the integrals related to $B$-mode power spectrum, the evaluation of the kSZ spectrum using the exact formulation is similarly time-complex.  However, unlike the approximated $B$-mode, the Limber approximated kSZ spectrum is rather consistent with the exact evaluation for the typical scales we are interested in ($\ell \geq 1000$). In Figure \ref{fig:limberksz512} we plot the kSZ spectrum evaluated using the exact formulation using Equation \eqref{eq:fullkSZ} and the kSZ spectrum evaluated from the Limber approximation using Equation \eqref{eq:ksz_lim}. We find that at all angular scales of our interest, the Limber approximated kSZ overlaps with the exact kSZ evaluation. We have checked and found that the match between the two power spectra is within numerical errors.

Based on the convergence of the two signals, we treat the Limber approximated kSZ to be a proxy for the exact evaluation with no visible effect on our bias studies.

\section{Estimating bias on the tensor-to-scalar ratio for realistic observing scenarios}\label{app:grn_restimates}

In order to simulate a more realistic observing scenario, we add Gaussian random noise with standard deviation dependent on the instrument noise specification and a zero mean to the mock power spectrum for each observing scenario. We investigate the bias in the same manner as presented in Table \ref{tab:r_estimates_nonoise} and \ref{tab:r_estimates3sigma_nonoise}, but with the modified mock power spectrum. The limits on all free parameters are shown in Table \ref{tab:paramconstfree}, with the bias estimates provided in the last column. For completeness, we also display the limits of derived model parameters in Table \ref{tab:paramconst}. We find that as a consequence of modification to the mock power spectrum, the recovery of $r$ for the model Template may slightly differ from the input $r$ but still represents the true recovery of $r$. The recovery of $r$ for model Template-$C^{\rm BB,reion}_{\ell}$ continues to represent a biased recovery. The crucial point is that the observed bias is consistent with the bias presented earlier in Table \ref{tab:r_estimates_nonoise} and Table \ref{tab:r_estimates3sigma_nonoise}. This consistency is expected, as the recovery of $r$ for both models is similarly affected when random noise is introduced into the mock spectrum and the statistical properties are obeyed. 

We also note that even after adding the additional signal of patchy $B$-mode through the model Template the constraints of both free and derived reionization parameters are surprisingly similar to the constraints obtained for the model Template - $\CBB{\rm reion}$. We find this trend consistent even with improved delensing ability. The probable cause for such a trend is that the constraints on reionization parameters are largely governed by error bars on Thomson scattering optical depth $\tau$ and the kSZ signal. 
\begin{table*} 
    \fontsize{8}{8}\selectfont
    \centering
    \caption{Parameter constraints for the free parameters, obtained from the MCMC analysis of the models Template and Template - $\CBB{\rm reion}$, are presented for different observatories and choices of reionization models. The constraints for the two mock values of $r$ used in this analysis, $10^{-3}$ and $5\times 10^{-4}$, are shown separately in the table. Their recovery is expressed in the form of $\cc{r^{\sigma_+}_{\sigma_-}}\times 10^3$. The bias estimate $\Delta r/\sigma_r$ is provided in the last column.}
    \begin{tabular}{lccccccc}
    \hline
    \multirow{2}{1em}{Observatory Case}     & \multirow{2}{1em}{Model} &  \multicolumn{5}{c}{68$\%$ limits} \\
    &&$\log\cc{M_{min,0}}$&$\log \cc{\zeta_{0}}$&$\alpha_\zeta$&$\alpha_M$&$10^3 \times r$& $\Delta r/\sigma$\\
    \hline
     & Input &9.73 & 1.57 & -2.01 & -2.06& $1$ &  \\
    \hline
    \multirow{2}{1em}{SO+}     & Template & \pc{9.67}{0.98}{0.42}& \pc{1.84}{0.43}{0.71} & \pc{-4.30}{2.89}{2.19} & $>-2.84$ & $<3.53$ & \multirow{2}{2em}{$-$}\\[0.075cm]
    & Template - $\CBB{\rm reion}$  & \pc{9.69}{0.95}{0.45} & \pc{1.85}{0.38}{0.75} & \pc{-4.25}{2.86}{2.27} & $>-2.82$ & $<3.54$ \\[0.12cm]
    \multirow{2}{1em}{LiteBIRD+}     & Template  & \pc{9.64}{0.95}{0.46} & \pc{1.77}{0.42}{0.67} & \pc{-3.95}{2.82}{2.15} & $>-2.94$ & \pc{0.859}{0.391}{0.601} & \multirow{2}{2em}{0.030}  \\[0.075cm]
    & Template - $\CBB{\rm reion}$  & \pc{9.68}{0.93}{0.46} & \pc{1.81}{0.41}{0.71} & \pc{-3.95}{2.88}{2.16}& $>-2.85$&\pc{0.874}{0.404}{0.560}  \\[0.12cm]
    \multirow{2}{1em}{CMBS4+}     & Template  & \pc{9.72}{0.79}{0.43} & \pc{1.65}{0.39}{0.53} & \pc{-2.25}{1.87}{0.99} & $>-3.24$ & \pc{0.986}{0.185}{0.185} & \multirow{2}{2em}{0.178}  \\[0.075cm]
    & Template - $\CBB{\rm reion}$  & \pc{9.73}{0.79}{0.45} & \pc{1.65}{0.39}{0.53} & \pc{-2.22}{1.59}{1.29}& $>-3.26$&\pc{1.019}{0.185}{0.186}  \\[0.12cm]
    \multirow{2}{1em}{PICO+}     & Template  & \pc{9.75}{0.79}{0.44} & \pc{1.66}{0.39}{0.54} & \pc{-2.08}{1.52}{1.37} & $>-3.25$ & \pc{0.948}{0.104}{0.105} & \multirow{2}{2em}{0.201} \\[0.075cm]
    & Template - $\CBB{\rm reion}$  & \pc{9.73}{0.80}{0.45} & \pc{1.64}{0.38}{0.55}& \pc{-2.02}{1.51}{1.24} & $>-3.78$&  \pc{0.969}{0.105}{0.106}  \\[0.12cm]
    \hline
     & Input &9.73 & 1.57 & -2.01 & -2.06 & 0.5  \\
    \hline
    \multirow{2}{1em}{CMBS4+}     & Template & \pc{9.75}{0.23}{0.37}& \pc{1.66}{0.38}{0.48} & \pc{-2.20}{1.50}{1.04} & $>-2.91$& \pc{0.497}{0.182}{0.182} & \multirow{2}{2em}{0.181}\\[0.075cm]
    & Template - $\CBB{\rm reion}$  &\pc{9.76}{0.70}{0.34} & \pc{1.66}{0.37}{0.45} & \pc{-2.12}{1.40}{0.92} & $>-2.72$ & \pc{0.530}{0.182}{0.181} \\[0.12cm]
    \multirow{2}{1em}{PICO+}     & Template & \pc{9.70}{0.84}{0.43}& \pc{1.63}{0.41}{0.54} & \pc{-2.10}{1.57}{1.28} & $>-3.24$& \pc{0.457}{0.096}{0.097} & \multirow{2}{2em}{0.227} \\[0.075cm]
    & Template - $\CBB{\rm reion}$  &\pc{9.74}{0.81}{0.44} & \pc{1.65}{0.38}{0.57} & \pc{-2.08}{1.57}{1.33} & $>-3.30$ & \pc{0.479}{0.097}{0.097} \\[0.12cm]
    \hline
    Delensing at $95\%$&&&&&&\\
    \multirow{2}{1em}{CMBS4+}     & Template & \pc{9.71}{0.81}{0.40}& \pc{1.64}{0.40}{0.52} & \pc{-2.29}{1.66}{1.19} & $>-3.29$& \pc{0.502}{0.122}{0.122} & \multirow{2}{2em}{0.246}\\[0.075cm]
    & Template - $\CBB{\rm reion}$  &\pc{9.73}{0.81}{0.40} & \pc{1.65}{0.40}{0.52} & \pc{-2.23}{1.68}{1.23} & $>-3.21$ & \pc{0.532}{0.121}{0.122} \\[0.12cm]
    
    \multirow{2}{1em}{PICO+}     & Template & \pc{9.73}{0.80}{0.41}& \pc{1.65}{0.40}{0.52} & \pc{-2.18}{1.60}{1.32} & $>-3.30$& \pc{0.469}{0.066}{0.064} & \multirow{2}{2em}{0.308}\\[0.075cm]
    & Template - $\CBB{\rm reion}$  &\pc{9.74}{0.79}{0.43} & \pc{1.65}{0.39}{0.46} & \pc{-2.11}{1.55}{1.25} & $>-3.24$ & \pc{0.489}{0.065}{0.074} \\[0.12cm]
    \hline
    Delensing at $100\%$&&&&&&\\
    \multirow{2}{1em}{CMBS4+}     & Template & \pc{9.72}{0.81}{0.42}& \pc{1.66}{0.40}{0.58} & \pc{-2.17}{1.59}{1.37} & $>-3.18$& \pc{0.493}{0.091}{0.091} & \multirow{2}{2em}{0.418}\\[0.075cm]
    & Template - $\CBB{\rm reion}$  &\pc{9.71}{0.83}{0.42} & \pc{1.65}{0.39}{0.55} & \pc{-2.27}{1.61}{1.29} & $>-3.25$ & \pc{0.531}{0.091}{0.091} \\[0.12cm]
    
    \multirow{2}{1em}{PICO+}     & Template & \pc{9.72}{0.87}{0.42}& \pc{1.64}{0.38}{0.58} & \pc{-2.03}{1.54}{1.28} & $>-3.20$& \pc{0.493}{0.016}{0.017} & \multirow{2}{2em}{1.588}\\[0.075cm]
    & Template - $\CBB{\rm reion}$  &\pc{9.82}{0.74}{0.36} & \pc{1.70}{0.41}{0.49} & \pc{-2.25}{1.55}{1.39} & $>-3.39$ & \pc{0.520}{0.017}{0.017} \\[0.12cm]
    \hline
     & Input: max-BB($3\sigma$)&$10.39$&$2.48$&$3.58$&$-0.76$&$1$\\
    \hline
    Delensing at $85\%$ & &&&\\ 
    \multirow{2}{1em}{CMBS4+}     & Template & \pc{9.76}{0.70}{0.39}& \pc{1.91}{0.42}{0.51} & \pc{0.92}{1.07}{1.88} & $>-3.23$& \pc{1.010}{0.186}{0.187} & \multirow{2}{2em}{0.320}\\[0.075cm]
    & Template - $\CBB{\rm reion}$  &\pc{9.72}{0.71}{0.40} & \pc{1.88}{0.41}{0.50} & \pc{0.85}{1.11}{1.67} & $>-3.23$ & \pc{1.070}{0.187}{0.187} \\[0.12cm]
    \multirow{2}{1em}{PICO+}     & Template & \pc{9.79}{0.69}{0.40}& \pc{1.93}{0.40}{0.54} & \pc{0.96}{1.09}{1.87} & $>-3.30$& \pc{0.956}{0.105}{0.105} & \multirow{2}{2em}{0.371}\\[0.075cm]
    & Template - $\CBB{\rm reion}$  &\pc{9.74}{0.68}{0.38} & \pc{1.88}{0.39}{0.50} & \pc{0.89}{1.06}{1.75} & $>-3.36$ & \pc{0.995}{0.105}{0.106} \\[0.12cm]
    \hline
    Delensing at $95\%$ & &&&\\ 
    \multirow{2}{1em}{CMBS4+}     & Template & \pc{9.77}{0.69}{0.44}& \pc{1.92}{0.42}{0.53} & \pc{0.87}{1.09}{1.85} & $>-3.36$& \pc{1.011}{0.131}{0.131} & \multirow{2}{2em}{0.534} \\[0.075cm]
    & Template - $\CBB{\rm reion}$  &\pc{9.76}{0.76}{0.42} & \pc{1.93}{0.43}{0.57} & \pc{1.07}{0.99}{1.88} & $>-3.03$ & \pc{1.081}{0.132}{0.131} \\[0.12cm]
    \multirow{2}{1em}{PICO+}     & Template & \pc{9.82}{0.65}{0.37}& \pc{1.95}{0.42}{0.48} & \pc{0.98}{1.11}{1.88} & $>-3.25$& \pc{0.976}{0.071}{0.071} & \multirow{2}{2em}{0.634}\\[0.075cm]
    & Template - $\CBB{\rm reion}$  &\pc{9.79}{0.67}{0.42} & \pc{1.93}{0.38}{0.54} & \pc{1.09}{1.06}{1.89} & $>-3.09$ & \pc{1.021}{0.071}{0.072} \\[0.12cm]
    \hline   
    & Input: max-BB($3\sigma$)&$10.39$&$2.48$&$3.58$&$-0.76$&$0.5$\\
    \hline
    Delensing at $85\%$ & &&&\\ 
    \multirow{2}{1em}{CMBS4+}     & Template & \pc{9.75}{0.72}{0.42}& \pc{1.91}{0.40}{0.56} & \pc{0.90}{1.07}{1.85} & $>-3.28$& \pc{0.510}{0.184}{0.185} & \multirow{2}{2em}{0.367}\\[0.075cm]
    & Template - $\CBB{\rm reion}$  &\pc{9.77}{0.68}{0.40} & \pc{1.92}{0.38}{0.53} & \pc{0.93}{1.08}{1.85} & $>-3.25$ & \pc{0.578}{0.185}{0.185} \\[0.12cm]
    \multirow{2}{1em}{PICO+}     & Template & \pc{9.81}{0.71}{0.39}& \pc{1.95}{0.42}{0.80} & \pc{1.09}{1.02}{2.05} & $>-3.13$& \pc{0.454}{0.098}{0.099} & \multirow{2}{2em}{0.426}\\[0.075cm]
    & Template - $\CBB{\rm reion}$  &\pc{9.80}{0.66}{0.43} & \pc{1.93}{0.36}{0.55} & \pc{1.04}{1.05}{1.92} & $>-3.21$ & \pc{0.496}{0.099}{0.099} \\[0.12cm]
    \hline
    Delensing at $95\%$ & &&&\\ 
    \multirow{2}{1em}{CMBS4+}     & Template & \pc{9.24}{0.71}{0.38}& \pc{1.90}{0.41}{0.52} & \pc{0.82}{1.06}{1.78} & $>-3.31$& \pc{0.514}{0.125}{0.125} & \multirow{2}{2em}{0.552} \\[0.075cm]
    & Template - $\CBB{\rm reion}$  &\pc{9.78}{0.70}{0.41} & \pc{1.93}{0.41}{0.54} & \pc{0.97}{1.00}{1.97} & $>-3.23$ & \pc{0.583}{0.125}{0.127} \\[0.12cm]
    \multirow{2}{1em}{PICO+}     & Template & \pc{9.85}{0.70}{0.34}& \pc{1.98}{0.50}{0.50} & \pc{1.14}{1.13}{1.96} & $>-2.98$& \pc{0.467}{0.067}{0.067} & \multirow{2}{2em}{0.726}\\[0.075cm]
    & Template - $\CBB{\rm reion}$  &\pc{9.79}{0.67}{0.41} & \pc{1.92}{0.39}{0.53} & \pc{0.96}{1.06}{1.92} & $>-3.39$ & \pc{0.516}{0.068}{0.068} \\[0.12cm]
    \hline
    \end{tabular}
    \label{tab:paramconstfree}
\end{table*}

\begin{table*} 
    \fontsize{8}{8}\selectfont
    \centering
    \caption{Same as Table \ref{tab:paramconstfree} but for the derived parameters}
    \begin{tabular}{lcccccccc}
    \hline
    \multirow{2}{1em}{Observatory Case}     & \multirow{2}{1em}{Model} &  \multicolumn{7}{c}{68$\%$ limits} \\
    &&$\tau$&$D^{\mathrm{kSZ}}_{\ell=3000}$&$D^{BB,\mathrm{reion}}_{\ell=200}$&$z_{25}$&$z_{50}$&$z_{75}$&$\Delta z$\\
    \hline
    $10^{-3}\times r= 1.0$ & Input &0.0540 & 3.00 & 7.03 & 8.09& 7.27 &  6.78 & 1.31 \\
    \hline
    \multirow{2}{1em}{SO+}     & Template & \pc{0.0579}{0.0054}{0.0042}& \pc{2.94}{0.09}{0.09} & \pc{6.39}{1.06}{1.54} & \pc{8.43}{0.46}{0.38}& \pc{7.72}{0.64}{0.46} & \pc{7.31}{0.79}{0.52}  & \pc{1.12}{0.24}{0.46}\\[0.075cm]
    & Template - $\CBB{\rm reion}$ &\pc{0.0579}{0.0056}{0.0043}& \pc{2.94}{0.09}{0.09} & \pc{6.43}{1.05}{1.56} & \pc{8.42}{0.48}{0.39}& \pc{7.71}{0.67}{0.47} & \pc{7.30}{0.81}{0.54}  & \pc{1.12}{0.24}{0.47}\\[0.12cm]
    \multirow{2}{1em}{LiteBIRD+}     & Template  & \pc{0.0572}{0.0055}{0.0043} & \pc{2.93}{0.09}{0.09} & \pc{6.46}{1.04}{1.49} & \pc{8.37}{0.52}{0.38} & \pc{7.63}{0.65}{0.49} & \pc{7.19}{0.81}{0.56}&\pc{1.18}{0.25}{0.49}  \\[0.075cm]
    & Template - $\CBB{\rm reion}$  & \pc{0.0572}{0.0054}{0.0045} & \pc{2.93}{0.09}{0.09} & \pc{6.51}{1.09}{1.49}& \pc{8.36}{0.47}{0.40}&\pc{7.62}{0.67}{0.49} & \pc{7.20}{0.81}{0.58} & \pc{1.16}{0.26}{0.48}   \\[0.12cm]
    \multirow{2}{1em}{CMBS4+}     & Template  & \pc{0.0538}{0.0019}{0.0020} & \pc{2.95}{0.07}{0.06} & \pc{7.28}{0.80}{1.26} & \pc{8.08}{0.81}{0.20} & \pc{7.22}{0.27}{0.25} & \pc{6.71}{0.43}{0.31}&\pc{1.37}{0.20}{0.44}  \\[0.075cm]
    & Template - $\CBB{\rm reion}$  & \pc{0.0537}{0.0019}{0.0019} & \pc{2.95}{0.07}{0.06} & \pc{7.31}{0.78}{1.28}& \pc{8.08}{0.18}{0.21}&\pc{7.21}{0.26}{0.26} & \pc{6.71}{0.41}{0.32} & \pc{1.37}{0.20}{0.43}   \\[0.12cm]
    \multirow{2}{1em}{PICO+}     & Template  & \pc{0.0534}{0.0018}{0.0018} & \pc{2.95}{0.07}{0.06} & \pc{7.40}{0.79}{1.32} & \pc{8.06}{0.16}{0.20} & \pc{7.18}{0.26}{0.24} & \pc{6.67}{0.41}{0.29}&\pc{1.38}{0.20}{0.43}  \\[0.075cm]
    & Template - $\CBB{\rm reion}$  & \pc{0.0532}{0.0018}{0.0018} & \pc{2.95}{0.07}{0.06} & \pc{7.39}{0.78}{1.33}& \pc{8.04}{0.18}{0.20}&\pc{7.16}{0.25}{0.20} & \pc{6.64}{0.41}{0.33} & \pc{1.40}{0.21}{0.44}   \\[0.12cm]
    \hline
    $ 10^{-3}\times r= 0.5$ & Input &0.0540 & 3.00 & 7.03 & 8.09& 7.27 &  6.78 & 1.31\\
    \hline
    \multirow{2}{1em}{CMBS4+}     & Template & \pc{0.0539}{0.0018}{0.0019}& \pc{2.95}{0.06}{0.07} & \pc{7.17}{0.69}{1.18} & \pc{8.09}{0.17}{0.20}& \pc{7.24}{0.28}{0.23} & \pc{6.75}{0.42}{0.28}  & \pc{1.33}{0.18}{0.41}\\[0.075cm]
    & Template - $\CBB{\rm reion}$  &\pc{0.0539}{0.0019}{0.0018} & \pc{2.95}{0.06}{0.07} & \pc{7.13}{0.62}{1.11} & \pc{8.09}{0.17}{0.19} & \pc{7.25}{0.29}{0.22} & \pc{6.77}{0.42}{0.48} & \pc{1.32}{0.17}{0.41}\\[0.12cm]
    \multirow{2}{1em}{PICO+}     & Template & \pc{0.0535}{0.0019}{0.0018}& \pc{2.95}{0.07}{0.06} & \pc{7.37}{0.82}{1.31} & \pc{8.07}{0.16}{0.21}& \pc{7.18}{0.27}{0.24} & \pc{6.66}{0.43}{0.29}  & \pc{1.41}{0.19}{0.46}\\[0.075cm]
    & Template - $\CBB{\rm reion}$  &\pc{0.0533}{0.0019}{0.0018} & \pc{2.95}{0.07}{0.06} & \pc{7.39}{0.81}{1.31} & \pc{8.05}{0.17}{0.21} & \pc{7.17}{0.27}{0.25} & \pc{6.65}{0.42}{0.31} & \pc{1.40}{0.20}{0.45}\\[0.12cm]
    \hline
    Delensing at $95\%$ & & & & & & & \\
    \multirow{2}{1em}{CMBS4+}     & Template & \pc{0.0538}{0.0020}{0.0020}& \pc{2.95}{0.07}{0.06} & \pc{7.24}{0.79}{1.24} & \pc{8.09}{0.20}{0.20}& \pc{7.23}{0.27}{0.26} & \pc{6.72}{0.43}{0.31}  & \pc{1.37}{0.18}{0.45}\\[0.075cm]
    & Template - $\CBB{\rm reion}$  &\pc{0.0538}{0.0020}{0.0020} & \pc{2.95}{0.07}{0.06} & \pc{7.28}{0.75}{1.28} & \pc{8.08}{0.19}{0.20} & \pc{7.22}{0.27}{0.26} & \pc{6.72}{0.42}{0.32} & \pc{1.36}{0.18}{0.43}\\[0.12cm]
    \multirow{2}{1em}{PICO+}     & Template & \pc{0.0536}{0.0019}{0.0019}& \pc{2.95}{0.06}{0.06} & \pc{7.33}{0.84}{1.27} & \pc{8.07}{0.18}{0.20}& \pc{7.20}{0.26}{0.24} & \pc{6.69}{0.40}{0.30}  & \pc{1.38}{0.20}{0.43}\\[0.075cm]
    & Template - $\CBB{\rm reion}$  &\pc{0.0534}{0.0019}{0.0019} & \pc{2.95}{0.06}{0.06} & \pc{7.36}{0.79}{1.29} & \pc{8.05}{0.18}{0.20} & \pc{7.18}{0.25}{0.26} & \pc{6.67}{0.40}{0.31} & \pc{1.39}{0.20}{0.43}\\[0.12cm]

    \hline
     Delensing at $100\%$ & & & & & & & \\
    \multirow{2}{1em}{CMBS4+}     & Template & \pc{0.0539}{0.0020}{0.0020}& \pc{2.96}{0.06}{0.06} & \pc{7.32}{0.74}{1.42} & \pc{8.10}{0.18}{0.21}& \pc{7.23}{0.28}{0.25} & \pc{6.73}{0.43}{0.31}  & \pc{1.37}{0.18}{0.45}\\[0.075cm]
    & Template - $\CBB{\rm reion}$  &\pc{0.0540}{0.0020}{0.0020} & \pc{2.95}{0.07}{0.06} & \pc{7.26}{0.78}{1.30} & \pc{8.10}{0.19}{0.21} & \pc{7.24}{0.24}{0.25} & \pc{6.74}{0.42}{0.32} & \pc{1.37}{0.17}{0.45}\\[0.12cm]
    
    \multirow{2}{1em}{PICO+}     & Template & \pc{0.0533}{0.0018}{0.0018}& \pc{2.95}{0.06}{0.06} & \pc{7.17}{0.25}{0.25} & \pc{8.05}{0.17}{0.20}& \pc{7.17}{0.25}{0.42} & \pc{6.64}{0.42}{0.24}  & \pc{1.41}{0.17}{0.46}\\[0.075cm]
    & Template - $\CBB{\rm reion}$  &\pc{0.0535}{0.0018}{0.0018} & \pc{2.95}{0.06}{0.06} & \pc{7.47}{0.82}{1.17} & \pc{8.04}{0.17}{0.18} & \pc{7.20}{0.26}{0.24} & \pc{6.71}{0.40}{0.29} & \pc{1.34}{0.16}{0.82}\\[0.12cm]
    \hline
    $10^3 \times r=1.0$& Input: max-BB ($3\sigma$) &0.0627 & 4.03 & 18.41 & 9.22 & 8.12 &  7.38 & 1.84\\
    \hline
    Delensing at $85\%$& & & & & & &  \\  
    \multirow{2}{1em}{CMBS4+}     & Template & \pc{0.0622}{0.0019}{0.0019}& \pc{3.99}{0.06}{0.06} & \pc{14.55}{0.98}{2.99} &
    \pc{9.25}{0.16}{0.22}  &
    \pc{7.89}{0.28}{0.24}& \pc{7.03}{0.46}{0.32} &  \pc{2.22}{0.23}{0.55}\\[0.075cm]
    & Template - $\CBB{\rm reion}$  &\pc{0.0621}{0.0019}{0.0019} & \pc{3.99}{0.06}{0.06} & \pc{14.35}{1.15}{2.75} & \pc{9.24}{0.16}{0.24} & \pc{7.86}{0.28}{0.24} & \pc{6.98}{0.47}{0.35} & \pc{2.26}{0.26}{0.58}\\[0.12cm]
    \multirow{2}{1em}{PICO+}     & Template & \pc{0.0619}{0.0018}{0.0018}& \pc{3.99}{0.07}{0.06} & \pc{14.75}{0.92}{3.06} &
    \pc{9.22}{0.16}{0.21}  &
    \pc{7.86}{0.29}{0.24}& \pc{7.00}{0.47}{0.32} &  \pc{2.22}{0.24}{0.54}\\[0.075cm]
    & Template - $\CBB{\rm reion}$  &\pc{0.0616}{0.0018}{0.0018} & \pc{3.99}{0.06}{0.06} & \pc{14.50}{1.08}{2.78} & \pc{9.20}{0.16}{0.22} & \pc{7.81}{0.27}{0.23} & \pc{6.92}{0.46}{0.33} & \pc{2.27}{0.25}{0.56}\\[0.12cm]
    \hline   
    Delensing at $95\%$ & & & & & & & \\ 
    \multirow{2}{1em}{CMBS4+}     & Template & \pc{0.0623}{0.0020}{0.0020}& \pc{3.99}{0.06}{0.06} & \pc{14.60}{1.07}{2.98} &
    \pc{9.25}{0.17}{0.22}  &
    \pc{7.89}{0.30}{0.26}& \pc{7.03}{0.48}{0.36} &  \pc{2.22}{0.27}{0.57}\\[0.075cm]
    & Template - $\CBB{\rm reion}$  &\pc{0.0622}{0.0020}{0.0020} & \pc{3.99}{0.07}{0.06} & \pc{14.73}{0.98}{3.41} & \pc{9.25}{0.16}{0.24} & \pc{7.89}{0.29}{0.24} & \pc{7.89}{0.29}{0.24} & \pc{2.23}{0.22}{0.57}\\[0.12cm]
    \multirow{2}{1em}{PICO+}     & Template & \pc{0.0620}{0.0018}{0.0018}& \pc{3.99}{0.06}{0.06} & \pc{14.81}{1.03}{2.98} &
    \pc{9.22}{0.15}{0.21}  &
    \pc{7.88}{0.28}{0.24}& \pc{7.02}{0.46}{0.31} &  \pc{2.20}{0.21}{0.53}\\[0.075cm]
    & Template - $\CBB{\rm reion}$  &\pc{0.0617}{0.0018}{0.0018} & \pc{3.99}{0.06}{0.06} & \pc{14.83}{0.88}{3.18} & \pc{9.20}{0.16}{0.21} & \pc{7.84}{0.28}{0.23} & \pc{6.97}{0.45}{0.31} & \pc{2.23}{0.22}{0.52}\\[0.12cm]
    \hline    
    $10^3 \times r=0.5$& Input: max-BB ($3\sigma$) &0.0627 & 4.03 & 18.41 & 9.22 & 8.12 &  7.38 & 1.84\\
    \hline
    Delensing at $85\%$& & & & & & &  \\  
    \multirow{2}{1em}{CMBS4+}     & Template & \pc{0.0623}{0.0019}{0.0019}& \pc{3.99}{0.06}{0.06} & \pc{14.52}{0.97}{3.03} &
    \pc{9.26}{0.16}{0.23}  &
    \pc{7.89}{0.30}{0.25}& \pc{7.03}{0.48}{0.35} &  \pc{2.23}{0.25}{0.57}\\[0.075cm]
    & Template - $\CBB{\rm reion}$  &\pc{0.0621}{0.0019}{0.0019} & \pc{3.99}{0.06}{0.06} & \pc{14.59}{0.94}{2.96} & \pc{9.23}{0.16}{0.22} & \pc{7.87}{0.28}{0.24} & \pc{7.01}{0.45}{0.34} & \pc{2.22}{0.24}{0.54}\\[0.12cm]
    \multirow{2}{1em}{PICO+}     & Template & \pc{0.0621}{0.0018}{0.0018}& \pc{3.99}{0.07}{0.06} & \pc{14.92}{0.96}{3.31} &
    \pc{9.22}{0.16}{0.21}  &
    \pc{7.87}{0.29}{0.24}& \pc{7.01}{0.46}{0.32} &  \pc{2.21}{0.22}{0.54}\\[0.075cm]
    & Template - $\CBB{\rm reion}$  &\pc{0.0616}{0.0018}{0.0018} & \pc{3.99}{0.06}{0.06} & \pc{14.84}{0.83}{3.14} & \pc{9.19}{0.17}{0.20} & \pc{7.83}{0.28}{0.24} & \pc{6.95}{0.45}{0.33} & \pc{2.24}{0.24}{0.53}\\[0.12cm]
    \hline   
    Delensing at $95\%$ & & & & & & & \\ 
    \multirow{2}{1em}{CMBS4+}     & Template & \pc{0.0623}{0.0020}{0.0020}& \pc{3.99}{0.06}{0.06} & \pc{14.40}{1.06}{2.81} &
    \pc{9.26}{0.16}{0.23}  &
    \pc{7.89}{0.29}{0.24}& \pc{7.03}{0.48}{0.35} &  \pc{2.23}{0.24}{0.58}\\[0.075cm]
    & Template - $\CBB{\rm reion}$  &\pc{0.0622}{0.0020}{0.0020} & \pc{3.99}{0.06}{0.06} & \pc{14.74}{0.89}{3.99} & \pc{9.24}{0.17}{0.22} & \pc{7.89}{0.29}{0.25} & \pc{7.89}{0.29}{0.25} & \pc{2.22}{0.23}{0.56}\\[0.12cm]
    \multirow{2}{1em}{PICO+}     & Template & \pc{0.0621}{0.0018}{0.0019}& \pc{3.99}{0.06}{0.06} & \pc{15.00}{1.10}{3.36} &
    \pc{9.22}{0.16}{0.21}  &
    \pc{7.90}{0.28}{0.24}& \pc{7.05}{0.45}{0.32} &  \pc{2.17}{0.19}{0.52}\\[0.075cm]
    & Template - $\CBB{\rm reion}$  &\pc{0.0618}{0.0018}{0.0019} & \pc{3.99}{0.06}{0.06} & \pc{14.76}{0.90}{3.03} & \pc{9.20}{0.17}{0.20} & \pc{7.84}{0.29}{0.29} & \pc{6.97}{0.46}{0.34} & \pc{2.24}{0.24}{0.55}\\[0.12cm]
    \hline    
    \end{tabular}
    \label{tab:paramconst}
\end{table*}


\bsp	
\label{lastpage}
\end{document}